\documentclass[12pt]{article}
\usepackage{jheppub_modified_green}
\usepackage{epsfig}
\usepackage{amsfonts}
\usepackage{amssymb}
\usepackage{enumitem}
\usepackage{yfonts}
\usepackage{amsmath}
\usepackage{amsthm}
\usepackage{subfig}
\usepackage{bm}
\usepackage{hyperref}
\usepackage{mathrsfs}
\usepackage{mathtools}
\usepackage{bbm}
\usepackage{cancel}
\usepackage{color,colortbl}
\usepackage{xcolor}
\usepackage{accents}
\usepackage{soul}
\usepackage{relsize}
\usepackage{float, slashed, graphicx, amssymb, amsmath}
\usepackage{shuffle}
\usepackage[normalem]{ulem}
\usepackage{tikz}
\usetikzlibrary{cd}

\allowdisplaybreaks 

\usetikzlibrary{external}             

\usepackage{tikz-feynman}
\tikzfeynmanset{compat=1.1.0}
\tikzfeynmanset{
graviton/.style={circle, draw=green!60, fill=green!5, very thick, minimum size=7mm}
}
\tikzfeynmanset{myblob/.style=
{/tikz/shape=rectangle,
/tikz/fill=red,
 /tikz/minimum size=0.3cm,
 } }
 \tikzfeynmanset{HV/.style=
{/tikz/shape=circle,
/tikz/fill={rgb:black,1;white,2},
 /tikz/minimum size=0.01cm,/tikz/inner sep=1.8pt
 } }
 \tikzfeynmanset{
codot/.style={/tikz/shape=circle,
/tikz/fill=red,/tikz/minimum size=0.2cm,/tikz/inner sep=.8pt}
}
\tikzfeynmanset{sb/.style=
{/tikz/shape=circle,
/tikz/fill=black,
 /tikz/minimum size=0.4cm,/tikz/inner sep=.8pt
 } }
  \tikzfeynmanset{GR/.style=
{/tikz/shape=ellipse,
/tikz/fill={rgb:black,1;white,2},
 /tikz/minimum size=0.3cm,
 } }
\tikzset{box/.pic={\filldraw[fill=black]  (0,0) circle (2.5pt);
				   \filldraw [fill=black] (0.5,0) circle (2.5pt);
			       \draw [line width=5pt] (0,0) -- (0.5,0);}}

\usepackage[T1]{fontenc} 

\newcommand{\Res}{ \text{Res}}

\makeatletter
\newcommand \UPlus {\mathop {\operator@font \uplus }\limits }
\makeatother
\makeatletter
\newcommand \Bigcup {\mathop {\operator@font \bigcup }\limits }
\makeatother
  \def\LabelNote#1{}
 \def\Label#1{\label{#1}%
  \smash{\hbox to0pt{\raise1ex\hbox{\tiny[#1]}\hss}}}
  
  \def\Cdot{{\cdot}}
  
\def\nn{\nonumber}

\newcommand{\widebar}{\overline}
\definecolor{applegreen}{rgb}{0.55, 0.71, 0.0}

\newcommand{\black}{\color{black}}

\newcommand{\white}{\color{white}}

\newcommand{\lan}{\langle}
\newcommand{\ran}{\rangle}

\newcommand{\cA}{\mathcal{A}}

\newcommand{\cM}{\mathcal{M}}
\newcommand{\cN}{\mathcal{N}}

\newcommand{\cO}{\mathcal{O}}

\newcommand{\eps}{\varepsilon}

\def\spa#1.#2{\left\langle#1\,#2\right\rangle}
\def\spb#1.#2{\left[#1\,#2\right]}

\def\be{\begin{align}}
\def\ee{\end{align}}
\def\bea{\begin{eqnarray}}
\def\eea{\end{eqnarray}}  
\allowdisplaybreaks

\newcommand{\npre}{\mathcal{N}}  

\newcommand{\commut}{\Gamma}

\newcommand{\cop}{\Delta}

\newcommand{\pole}[1]{\left({1\over #1} \right)}

\DeclareMathOperator{\cC}{\mathcal{C}}

\newcommand{\Id}{\mathbbm{I}}
\def\one{\mbox{1 \kern-.59em {\rm l}}}

\usepackage{tikz}

\title{Amplitudes, Hopf algebras and the colour-kinematics duality}
\author{Andreas Brandhuber$^{a,b}$,}
\author{Graham R.~Brown$^{a,b}$,}
\author{Gang Chen$^{a}$,}
\author{Joshua Gowdy$^{a,b}$,}
\author{\\Gabriele Travaglini$^{a,b}$}
\author{and Congkao Wen$^{a}$}
\affiliation{(a)~Centre for Theoretical Physics, Department of Physics and Astronomy, \\
Queen Mary University of London, Mile End Road, London E1 4NS, United Kingdom}

\affiliation{(b)~Kavli Institute for Theoretical Physics, University of California, Santa Barbara, 
\\CA~93106, USA}

\emailAdd{a.brandhuber@qmul.ac.uk}
\emailAdd{graham.brown@qmul.ac.uk}
\emailAdd{g.chen@qmul.ac.uk}
\emailAdd{j.gowdy@qmul.ac.uk}
\emailAdd{g.travaglini@qmul.ac.uk}
\emailAdd{c.wen@qmul.ac.uk}

\begin{document} 
\begin{flushright}
	QMUL-PH-22-18\\
	SAGEX-22-27\\
\end{flushright}


\abstract{
It was recently proposed that the kinematic algebra featuring in the colour-kinematics  duality for scattering amplitudes in heavy-mass effective field theory (HEFT) and Yang-Mills theory is a quasi-shuffle Hopf algebra. The associated fusion product determines the structure of the Bern-Carrasco-Johansson (BCJ) numerators, which  are manifestly gauge invariant and  with poles corresponding to heavy-particle exchange. In this work we explore the deep connections between the quasi-shuffle algebra and general physical properties of the scattering amplitudes. First, after proving the double-copy form for gravitational HEFT amplitudes, we show that the coproducts of the kinematic algebra are in correspondence with factorisations of  BCJ numerators on massive poles. We then study an extension of the standard quasi-shuffle Hopf algebra to a non-abelian version describing BCJ numerators with all possible gluon orderings. This is achieved by tensoring the original algebra with a particular Hopf algebra of orderings. In this extended version, a specific choice of the coproduct in the algebra of orderings leads to an  antipode in the resulting Hopf algebra that has the interpretation of reversing the gluons' order within each BCJ numerator.

}

\maketitle
\flushbottom

\newpage

\section{Introduction}
The colour-kinematics duality and double copy \cite{Bern:2008qj,Bern:2010ue,Bern:2019prr} play an important role in the modern approach to  scattering amplitudes and gravitational physics.
Inspired by Kawai-Lewellen-Tye (KLT) relations  \cite{Kawai:1985xq} for string amplitudes, these remarkable properties  were originally proposed for tree-level amplitudes in pure Yang-Mills theory and gravity, and  extended to amplitudes in a variety of other  theories \cite{Bargheer:2012gv,Broedel:2012rc,Chiodaroli:2013upa,Johansson:2014zca,Chiodaroli:2014xia,Chiodaroli:2015rdg,Johansson:2017srf,Chiodaroli:2018dbu,Chen:2013fya,Carrasco:2016ldy,Mafra:2016mcc,Carrasco:2016ygv, Cheung:2016prv, Cheung:2017ems,Cheung:2017yef,Bonnefoy:2021qgu,Carrasco:2019yyn,Carrasco:2021ptp,Chi:2021mio,Menezes:2021dyp}, to loop level \cite{Bern:2012uf,Bjerrum-Bohr:2013iza,Bern:2013yya,Nohle:2013bfa,Mogull:2015adi,He:2017spx,Johansson:2017bfl,Boels:2012ew,Yang:2016ear,Lin:2020dyj} and even  to exact classical solutions    \cite{Monteiro:2014cda, Luna:2018dpt,Armstrong-Williams:2022apo}. 
Recently the underlying algebra and  geometry of the colour-kinematics duality and double copy have been studied from several  different viewpoints
\cite{Monteiro:2011pc,BjerrumBohr:2012mg,Fu:2016plh,Cheung:2016prv,Cheung:2017yef, Reiterer:2019dys,Tolotti:2013caa,Mizera:2019blq,Borsten:2020zgj,Borsten:2020xbt,Borsten:2021hua,Ferrero:2020vww,Cheung:2021zvb,Cheung:2022vnd,White:2020sfn,Guevara:2021yud,Cohen:2022uuw,Diaz-Jaramillo:2021wtl,Chen:2019ywi,Chen:2021chy,Brandhuber:2021kpo,Brandhuber:2021bsf}, leading to an in-depth exploration of gravitational physics, gauge field theory and their 
relationships.

 In this paper we mainly focus on the algebraic aspects of the colour-kinematics duality.
The first critical milestone is to reveal the kinematic algebra and construct all the  duality-satisfying numerators, known as Bern-Carrasco-Johansson (BCJ) numerators, from an algebraic product. In \cite{Brandhuber:2021bsf}, and building on \cite{Chen:2019ywi,Chen:2021chy,Brandhuber:2021kpo}, four of the present authors and Johansson proposed a kinematic algebra that underlies the colour-kinematics duality in a heavy-mass effective theory (HEFT) and, using a decoupling limit,  in Yang-Mills (YM) theory.
 Remarkably, it was found that  the BCJ numerators for tree-level amplitudes of two heavy particles and any number of gluons can be constructed from a quasi-shuffle Hopf algebra, which is well studied in the context of combinatorial Hopf algebras of shuffles and quasi-shuffles \cite{hoffman2000quasi,Blumlein:2003gb,aguiar2010monoidal,hoffman2017quasi,fauvet2017hopf}. 
 
  The HEFT BCJ numerators obtained in our approach  enjoy  several  important properties: 

 \noindent
 {\bf 1.}~They  are manifestly gauge invariant. 
 
 \noindent
 {\bf 2.}~They are local for  massless gluons, but  contain  poles corresponding to the propagation of the heavy particles. 

\noindent
{\bf 3.}~Higher-point numerators factorise into products of lower-point ones at the singularities of the massive poles.

 \noindent
 {\bf 4.}~By decoupling the  two heavy particles  through a factorisation limit, one obtains  BCJ numerators for tree-level amplitudes in pure YM theory, which therefore  are also constructed using the same quasi-shuffle Hopf  algebra.

This paper sharpens  our understanding of  the kinematic  algebra and its applications to the understanding of the colour-kinematics duality in HEFT and YM theory.  
Our goal is twofold: first,  we wish to link various mathematical aspects of the newly discovered quasi-shuffle Hopf algebra \cite{Brandhuber:2021bsf} back to the physical properties of scattering amplitudes; and second, we  introduce new mathematical structures and extend further the quasi-shuffle Hopf algebra to incorporate certain general properties of scattering amplitudes, which are otherwise difficult to describe using the standard quasi-shuffle algebra. These two directions  go hand-in-hand.

 \black 
Specifically, for better describing the ordering of particles in the scattering amplitudes we are led to a non-abelian extension of the standard quasi-shuffle Hopf algebra, and  important structures of the algebra such as the coproduct and the antipode act non-trivially on the BCJ numerators. As we will show, the action of the coproduct is interpreted as a factorisation limit of the BCJ numerators on the massive poles, whereas the antipode reverses the ordering of the particles entering the numerators. 

There are several reasons to pursue these goals. 
One stems from  the importance  of exploring the kinematic algebra from a mathematical viewpoint. In addition, one may recall that the gravitational amplitudes obtained after a double copy describe black hole scattering and gravitational wave emission.  The resulting expressions in the approach described in this paper are extremely compact, and have led to one of the most efficient ways of studying black hole scattering \cite{Brandhuber:2021eyq}.

The rest of the paper is organised as follows. In Section~\ref{eq:HEFT-bcj}, we  review the novel colour-kinematics duality and general properties of  tree-level amplitudes in HEFT, including the decoupling limit leading to YM amplitudes. We  also prove the double copy formula for the gravitational amplitudes involving heavy particles using  KLT relations.  
In Section~\ref{sec:3}, we introduce the concept of the pre-numerators in HEFT and their construction from abstract algebraic generators.  These generators obey a quasi-shuffle Hopf algebra, and the pre-numerators are obtained using a map from the generators to functions of momenta and polarisation vectors. The BCJ numerators are directly related to the pre-numerators in a simple manner through the so-called nested commutators. 
Section~\ref{sec:Hopf-algebra} aims to study the kinematic Hopf algebra for the BCJ numerators in HEFT and to understand the physical interpretation of various mathematical operations in the Hopf algebra. In particular, the coproduct is related to the factorisation of the pre-numerators. In 
Section~\ref{sec:non-abelian}, we introduce the non-abelian version of the quasi-shuffle Hopf algebra. This is realised by extending the algebraic generators to include the ordering of external particles. We show that it is this version of the kinematic algebra that directly  describes the BCJ numerators in an algebraic fashion   in  the HEFT.
Section~\ref{sec:factorisation2} is concerned with the factorisation behaviour of BCJ numerators and tree-level amplitudes in HEFT and their connections with coproducts, using the non-abelian version of the quasi-shuffle Hopf algebra. Finally, we conclude in Section~\ref{sec: conclusion} and comment on several directions for further research. Three appendices complete the paper. In 
Appendix~\ref{app: RecursiveDefs} we provide  a recursive definition of the quasi-shuffle product, while in  
Appendix~\ref{app:B} 
we discuss a connection between shuffle and quasi-shuffle algebras. Finally, in Appendix~\ref{ap: antipodePreNum} we give an alternative definition of the pre-numerator from a different mapping rule which ensures that the pre-numerators themselves are symmetric under the action of the antipode.

\section{Colour-kinematics duality and double copy  in HEFT}
\label{eq:HEFT-bcj}

\subsection{General setup}

We will be interested in amplitudes involving two massive scalars with mass $m$ and $n{-}2$ gluons/gravitons, in the heavy-mass limit~\cite{Georgi:1990um, Luke:1992cs, Neubert:1993mb, Manohar:2000dt,Damgaard:2019lfh}. More precisely, we will define the HEFT amplitude to be the piece of the gluon or graviton amplitude which is of order $m$ or $m^2$, respectively%
\footnote{Strictly speaking, gravity amplitudes contain additional {\it contact terms} of order $m^3$ and higher~\cite{Brandhuber:2021eyq} in the heavy-mass limit. These are given by products of lower-point HEFT amplitudes and delta functions, and vanish on generic kinematic configurations. In the following we will ignore such contributions, which is equivalent to dropping all Feynman $i \epsilon$'s.}. For Yang-Mills-scalar amplitudes   the HEFT amplitudes are simply obtained  by taking the leading term in the heavy-mass limit. For instance,  
the colour-ordered two-scalar/two-gluon amplitude is given by%
\footnote{We have omitted the overall coupling dependence.} 
\begin{equation}
        \mathcal{A}(1,2,3,4)=\frac{2 p_3\Cdot F_{1}\Cdot F_{2}\Cdot p_3}{s_{12} \big(s_{13}-m^2\big)}\,,
\end{equation}
where particles $1,2$ are gluons,  $3,4$ are adjoint massive scalars.    $F_i^{\mu\nu}{:=} p_i^\mu \eps_i^\nu - p_i^\nu \eps_i^\mu$ denotes a linearised field strength, and $s_{ij}{:=}(p_i+p_j)^2$. To take the heavy-mass limit we write the momentum of the first and second scalar (in a convention where all particles are outgoing) as $p_3^\mu= -m v^\mu$ and  $p_4^\mu= m v^\mu+q^\mu$, respectively, where $v^\mu$ is the velocity of the heavy particle and  $q^\mu=p_{1}^{\mu}+p_{2}^{\mu}$ is the sum of the gluon momenta.
Then  we can take the leading term in the large-$m$ expansion, to yield
\begin{equation}
     A(12,v)=-\frac{m}{s_{12}}\frac{v\Cdot F_{1}\Cdot F_{2}\Cdot v}{v\Cdot p_{1}}\,.
\end{equation}
In the above we have replaced the two scalar labels $3,4$ with a single label for the velocity $v$. For $n$-point HEFT amplitudes we will also make the same replacement, with legs $n{-}1$ and $n$ always labelling the heavy scalars,
\begin{align}
     &\mathcal{A}(1,2,\ldots,n{-}1,n)\xrightarrow[]{\rm HEFT} A(12\ldots n{-}2,v)\,.
\end{align}
In the same manner as above, we can take the heavy-mass limit of a gravitational amplitude $\cM(1,2,\ldots,n{-}1,n)$ involving two massive scalars and extract the term which scales like $m^2$ giving the gravitational HEFT amplitude $M(12\ldots n{-}2,v)$
\begin{align}
     &\mathcal{M}(1,2,\ldots,n{-}1,n)\xrightarrow[]{\rm HEFT} M(12\ldots n{-}2,v)\, .
\end{align}

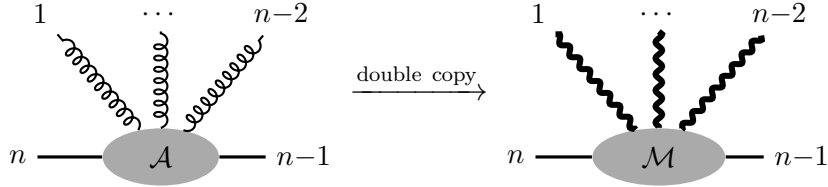
\begin{align}\nn
\begin{tikzpicture}[baseline={([yshift=-0.8ex]current bounding box.center)}]\tikzstyle{every node}=[font=\small]    
   \begin{feynman}
    \vertex (a)[]{$n{-}1$};
     \vertex[left=1.9cm of a] (a0)[GR]{$~~\rm \cA~~$};
      \vertex[left=1.9cm of a0] (am1){$n$};
    \vertex[above=1.9cm of a0] (b0){$\cdots$};
     \vertex[left=1.6cm of b0] (bm1){$1$};
      \vertex[right=1.6cm of b0] (b1){${n{-}2}$};
   	 \diagram*{(a)--[very thick](a0),(am1)--[very thick](a0), (a0) -- [thick,gluon] (b0),(a0) -- [thick,gluon] (bm1),(a0) -- [thick,gluon] (b1)};
    \end{feynman}  
  \end{tikzpicture}\xrightarrow[]{\rm double ~ copy} \begin{tikzpicture}[baseline={([yshift=-0.8ex]current bounding box.center)}]
  \tikzstyle{every node}=[font=\small]    
   \begin{feynman}
    \vertex (a)[]{$n{-}1$};
     \vertex[left=1.9cm of a] (a0)[GR]{$~~\rm \cM~~$};
      \vertex[left=1.9cm of a0] (am1){$n$};
    \vertex[above=1.9cm of a0] (b0){$\cdots$};
     \vertex[left=1.6cm of b0] (bm1){$1$};
      \vertex[right=1.6cm of b0] (b1){${n{-}2}$};
   	 \diagram*{(a)--[very thick](a0),(am1)--[very thick](a0), (a0) -- [photon,line width=0.8mm] (b0),(a0) -- [line width=0.8mm,photon] (bm1),(a0) -- [line width=0.8mm,photon] (b1)};
    \end{feynman}  
  \end{tikzpicture}
  \end{align}
Note that gluons (gravitons) in YM (gravity) amplitudes $\mathcal{A}$ ($\mathcal{M}$) are labelled as $1, \ldots, n{-}2$ as in the figure above. 
We also quote  the useful relation 
\begin{equation}
    v\Cdot p_{12\ldots n-2}=0+\cO(m^{-1})\, , 
\end{equation}
which follows from  momentum conservation and on-shell conditions, where we have defined $p_{12\ldots m}{:=}\sum_{k=1}^m p_k $. From now on we will drop explicit factors of $m$ and $m^2$ since HEFT amplitudes always scale homogeneously in the mass.

A novel colour-kinematic duality and double copy for HEFT amplitudes  were recently proposed in \cite{Brandhuber:2021kpo, Brandhuber:2021bsf}.
The colour-ordered YM tree amplitudes with two heavy particles and $n{-}2$ gluons are found to take the following form
\begin{align}
\label{eq:newDC}
	A(12\ldots n{-}2,v)&\, =\, \sum_{\commut \in \rho} {\npre(\commut,v)\over d_\commut}\, ,
\end{align}
where $\rho$ denotes all ordered nested commutators of the  labels $\{ 1, \ldots \,  , n{-}2 \}$. 
Each term in the sum is in one-to-one correspondence with a cubic graph with gluon labels following the colour ordering of the amplitude,  and 
$d_\commut$ is the product of inverse scalar propagators of massless particles associated with that  cubic graph.
 $\npre(\commut,v)$ represents the HEFT  BCJ numerator for the cubic graph $\Gamma$, whose properties and explicit construction will be studied in detail in this paper. Note that the number of terms in the sum in \eqref{eq:newDC} is given by the Catalan number $C_{n-3}$ (for  $n$-gluon amplitudes in YM,  this number becomes $C_{n-2}$).

As explained in e.g.~\cite{Chen:2021chy}, every nested commutator structure has an associated cubic graph, where the vertices of the graph correspond to commutators and the external legs correspond to gluons or gravitons. For gluons we only consider graphs where the external legs follow the colour ordering of the amplitude. As an example, for the set $\{1,2,3\}$, we have    $\rho{=}\{[[1,2],3],\,[1,[2,3]]\}$, and for the set $\{1,2,3,4\}$ we have $\rho{=}\{[[1,2],[3,4]],\,[[1,2],3],4],\,[1,[2,[3,4]]],\,[[1,[2,3]],4],\,[1,[[2,3],4]]\}$. 
The first nested commutator in each list can be represented pictorially as
\begin{equation}\label{nested}
   \begin{tikzpicture}[baseline={([yshift=-0.8ex]current bounding box.center)}]\tikzstyle{every node}=[font=\small]    
   \begin{feynman}
    \vertex (a)[myblob]{};
     \vertex [above=0.3cm of a](b)[dot]{};
     \vertex [left=0.6cm of b](c);
     \vertex [left=0.22cm of b](c23);
     \vertex [above=0.14cm of c23](v23)[dot]{};
    \vertex [above=.4cm of c](j1){$1$};
    \vertex [right=.7cm of j1](j2){$2$};
    \vertex [right=0.5cm of j2](j3){$3$};
   	 \diagram*{(a) -- [thick] (b),(b) -- [thick] (j1),(v23) -- [thick] (j2),(b)--[thick](j3)};
    \end{feynman}  
  \end{tikzpicture}\!\!\quad \longrightarrow \quad  \cN([[1,2], 3],v)\,,
  \quad 
     \begin{tikzpicture}[baseline={([yshift=-0.8ex]current bounding box.center)}]\tikzstyle{every node}=[font=\small]    
   \begin{feynman}
    \vertex (a)[myblob]{};
     \vertex [above=0.3cm of a](b)[dot]{};
     \vertex [left=0.8cm of b](c);
     \vertex [left=0.3cm of b](c12);
     \vertex [right=0.3cm of b](c34);
     \vertex [above=0.17cm of c12](v12)[dot]{};
     \vertex [above=0.17cm of c34](v34)[dot]{};
    \vertex [above=.4cm of c](j1){$1$};
    \vertex [right=.7cm of j1](j2){};
    \vertex [right=0.5cm of j2](j3){};
    \vertex [right=0.35cm of j3](j4){};
    \vertex [below=0.08cm of j4](k4){};
    \vertex [below=0.05cm of j2](k2){};
    \vertex [below=0.05cm of j3](w3){};
    \vertex [left=0.36cm of w3](k3){};
    \vertex [right=0.6cm of j1](l2){$2$};
    \vertex [right=1.0cm of j1](l3){$3$};
    \vertex [right=1.5cm of j1](l4){$4$};
   	 \diagram*{(a) -- [thick] (b),(b) -- [thick] (j1),(v12) -- [thick] (k2),(b)--[thick](k4),(v34)--[thick](k3)};
    \end{feynman}  
  \end{tikzpicture}\!\!\quad \longrightarrow \quad  \cN([[1,2], [3,4]],v)\,,
\end{equation}
where the red box represents the two heavy scalars. 
For these particular graphs,  we also have  $d_\commut = s_{12} s_{123}$ and 
$d_\commut = s_{12} s_{34} s_{1234}$, respectively. 

We will show in the next section that 
once the gluon amplitudes are recast in  colour-kinematics form \eqref{eq:newDC},  then the double-copy procedure will allow us to directly write down the tree-level amplitude of $n{-}2$ gravitons coupled to two heavy particles~as 
\begin{align}
	\label{eq:newDC-bis}
	M(12\ldots n{-}2,v)&\, =\, \sum_{\commut\in \widetilde{\rho}} {\big[\npre(\commut,v)\big]^2 \over d_\commut}\, .
\end{align}
There is no colour ordering for gravitons, therefore $\widetilde{\rho}$ denotes all un-ordered nested commutators of $\{ 1, \ldots \,  , n{-}2 \}$, which are labels for the external gravitons. For example,  in the case of three gravitons we have 
 $\tilde{\rho}=\{[[1,2],3],[[1,3],2],[1,[2,3]]\}$.
 In general, the total number of terms of un-ordered nested commutators in
\eqref{eq:newDC-bis} is $(2n-7)!!$. Each term in the sum is in one-to-one correspondence with a cubic graph.\black

The BCJ numerators are manifestly gauge invariant and contain only physical 
heavy-mass propagators, which are linear. Thus, by construction,  gluon HEFT amplitudes  contain only single massive poles, while  graviton HEFT amplitudes contain only double massive poles. The latter propery  is entirely non-obvious  upon  simply   expanding  graviton amplitudes in the full theory in the large-mass limit. 
Moreover, the antisymmetric property for a single graph: 
\begin{align}
\label{anti-prop}
\begin{tikzpicture}[baseline={([yshift=-0.8ex]current bounding box.center)}]\, \tikzstyle{every node}=[font=\small]    
   \begin{feynman}
    \vertex (a)[HV]{$~~~~$};
     \vertex [above=0.6cm of a](b)[dot]{};
     \vertex [left=1.6cm of b](c);
     \vertex [left=0.82cm of b](cL);
     \vertex [above=0.8cm of cL](vL)[HV]{$L$};
     \vertex [right=1.64cm of vL](vR)[HV]{$R$};
     \vertex [above=0.6cm of vL](t1){};
     \vertex [above=0.6cm of vR](t2){};
    \vertex [above=1.8cm of c](j1){};
    \vertex [right=1.2cm of j1](j2){};
     \vertex [right=1.1cm of j2](j3){};
    \vertex [right=0.8cm of j3](j4){};
   	 \diagram*{(a) -- [thick] (b),(b)--[thick](vL) -- [thick] (j1),(vL) -- [thick] (j2),(vR)--[thick](j3),(b)--[thick](vR)--[thick](j4)};
    \end{feynman}  \, ,
  \end{tikzpicture}  =- \begin{tikzpicture}[baseline={([yshift=-0.8ex]current bounding box.center)}]\, \tikzstyle{every node}=[font=\small]    
   \begin{feynman}
    \vertex (a)[HV]{$~~~~$};
     \vertex [above=0.6cm of a](b)[dot]{};
     \vertex [left=1.6cm of b](c);
     \vertex [left=0.82cm of b](cL);
     \vertex [above=0.8cm of cL](vL)[HV]{$R$};
     \vertex [right=1.64cm of vL](vR)[HV]{$L$};
     \vertex [above=0.6cm of vL](t1){};
     \vertex [above=0.6cm of vR](t2){};
    \vertex [above=1.8cm of c](j1){};
    \vertex [right=1.2cm of j1](j2){};
     \vertex [right=1.1cm of j2](j3){};
    \vertex [right=0.8cm of j3](j4){};
   	 \diagram*{(a) -- [thick] (b),(b)--[thick](vL) -- [thick] (j1),(vL) -- [thick] (j2),(vR)--[thick](j3),(b)--[thick](vR)--[thick](j4)};
    \end{feynman}  \, ,
  \end{tikzpicture}
  \end{align}
  and the Jacobi identity  for a  triplet of graphs:
  \begin{align}
  \label{Jaco-prop}
   \begin{tikzpicture}[baseline={([yshift=-0.8ex]current bounding box.center)}]\, \tikzstyle{every node}=[font=\small]    
   \begin{feynman}
    \vertex (a)[HV]{$~~~~$};
     \vertex [above=0.6cm of a](b)[dot]{};
     \vertex [left=0.45cm of b](c23);
     \vertex [above=0.5cm of c23](v23)[dot]{};
     \vertex [left=1.4cm of b](c);
     \vertex [left=1.02cm of b](cL);
     \vertex [above=1.0cm of cL](vL)[HV]{$A$};
     \vertex [right=2.04cm of vL](vR)[HV]{$C$};
       \vertex [right=1.1cm of vL](vm)[HV]{$B$};
    \vertex [above=2.0cm of c](j1){};
    \vertex [right=0.8cm of j1](j2){};
       \vertex [right=0.4cm of j2](j5){};
    \vertex [right=0.8cm of j5](j6){};
     \vertex [right=1.4cm of j2](j3){};
    \vertex [right=0.7cm of j3](j4){};
   	 \diagram*{(a) --[thick] (b),(b)--[thick](vL) -- [thick] (j1),(vL) -- [thick] (j2),(vR)--[thick](j3),(b)--[thick](vR)--[thick](j4),(vm)--[thick](j6),(vm)--[thick](j5),(v23)--[thick](vm)};
    \end{feynman}  \, ,
  \end{tikzpicture} -
 \begin{tikzpicture}[baseline={([yshift=-0.8ex]current bounding box.center)}]\, \tikzstyle{every node}=[font=\small]    
   \begin{feynman}
    \vertex (a)[HV]{$~~~~$};
     \vertex [above=0.6cm of a](b)[dot]{};
     \vertex [left=0.45cm of b](c23);
     \vertex [above=0.5cm of c23](v23)[dot]{};
     \vertex [left=1.4cm of b](c);
     \vertex [left=1.02cm of b](cL);
     \vertex [above=1.0cm of cL](vL)[HV]{$A$};
     \vertex [right=2.04cm of vL](vR)[HV]{$B$};
       \vertex [right=1.1cm of vL](vm)[HV]{$C$};
    \vertex [above=2.0cm of c](j1){};
    \vertex [right=0.8cm of j1](j2){};
       \vertex [right=0.4cm of j2](j5){};
    \vertex [right=0.8cm of j5](j6){};
     \vertex [right=1.4cm of j2](j3){};
    \vertex [right=0.7cm of j3](j4){};
   	 \diagram*{(a) --[thick] (b),(b)--[thick](vL) -- [thick] (j1),(vL) -- [thick] (j2),(vR)--[thick](j3),(b)--[thick](vR)--[thick](j4),(vm)--[thick](j6),(vm)--[thick](j5),(v23)--[thick](vm)};
    \end{feynman}  \, ,
  \end{tikzpicture} =\begin{tikzpicture}[baseline={([yshift=-0.8ex]current bounding box.center)}]\, \tikzstyle{every node}=[font=\small]    
   \begin{feynman}
    \vertex (a)[HV]{$~~~~$};
     \vertex [above=0.6cm of a](b)[dot]{};
     \vertex [right=0.45cm of b](c23);
     \vertex [above=0.5cm of c23](v23)[dot]{};
     \vertex [left=1.4cm of b](c);
     \vertex [left=1.02cm of b](cL);
     \vertex [above=1.0cm of cL](vL)[HV]{$A$};
     \vertex [right=2.04cm of vL](vR)[HV]{$C$};
       \vertex [right=1.1cm of vL](vm)[HV]{$B$};
    \vertex [above=2.0cm of c](j1){};
    \vertex [right=0.8cm of j1](j2){};
       \vertex [right=0.4cm of j2](j5){};
    \vertex [right=0.8cm of j5](j6){};
     \vertex [right=1.4cm of j2](j3){};
    \vertex [right=0.7cm of j3](j4){};
   	 \diagram*{(a) --[thick] (b),(b)--[thick](vL) -- [thick] (j1),(vL) -- [thick] (j2),(vR)--[thick](j3),(b)--[thick](vR)--[thick](j4),(vm)--[thick](j6),(vm)--[thick](j5),(v23)--[thick](vm)};
    \end{feynman}  \, ,
  \end{tikzpicture}
  \end{align}
  hold automatically. We also note the following two important relations for  colour-ordered gluon amplitudes:
\begin{align}\label{eq: HEFTampreversal}
  & \textit{U(1) decoupling relation:}    \sum_{\sigma\in i\shuffle\{1,\ldots,i-1,i+1,\ldots,n-2\}} A(\sigma,v)=0\, , \\
 & \textit{Reflection relation:} 
 \label{eq:reflrel}
 \ \  \ \  \ \ \ \ \ A(12\ldots n{-}2, v ) = (-1)^{n{-}1}A(n{-}2\ldots 21, v)\, .
\end{align}
For example, we have
\begin{align}
\begin{split}
	A(123,v)+A(213,v)+A(231,v)&=0\, , \\
		A(123,v)-A(321,v)&=0\, .
		\end{split}
\end{align}
\black
Finally,  it is straightforward  to obtain  pure Yang-Mills amplitudes or BCJ numerators from  the HEFT numerators by taking the decoupling limit \cite{Brandhuber:2021bsf}
\begin{align}
\label{dec-limit} 
v\to \eps_{n-1}\, , 
\qquad 
(p_{1}+\cdots +p_{n-2})^2\to 0\, , 
\end{align}
This limit  maps an $n$-point HEFT amplitude (with two scalars and $n{-}2$ gluons) into an $(n{-}1)$-point YM amplitude, with the two scalars replaced by gluon $n{-}1$. 

\subsection{Proof of the HEFT double copy}

The particular representation of  HEFT amplitudes given in \eqref{eq:newDC} was proved in \cite{Brandhuber:2021bsf} by understanding the factorisation behaviour of the BCJ numerators. The main goal of this  section is to  prove the double-copy formula \eqref{eq:newDC-bis}. To do so, we start from  the KLT relation for  the full theory  \cite{Bjerrum-Bohr:2010pnr} 
\begin{align}
   {\cal M}(1,2, \ldots, n{-}1,n)=-\sum_{\alpha,\beta\in S_{n-3}}\mathcal{S}(1\alpha  |1\beta  ){\cal A}(1,\alpha,n{-}1,n){\cal A}(1,\beta, n, n{-}1)\, 
\, , \end{align}
 where 
$\mathcal{S}(1\alpha |1\beta)$ is the field theory KLT kernel which is an $(n{-}3)! \times (n{-}3)!$ matrix, and the sum is over the $(n{-}3)!$ permutations of $(2, \ldots , n{-}2)$. 
In the heavy-mass limit for the particles  $n{-}1$ and $n$, we see that swapping $n{-}1$ and $n$ amounts to sending $v{\to} -v$, and the HEFT amplitudes are odd under this transformation, 
\begin{align}\label{eq: vtominusv}
\begin{split}
    {\cal A}(1,\alpha, n{-}1, n) &\stackrel{\rm HEFT}{\longrightarrow}A(1 \alpha, v)\ ,  \\
    {\cal A}(1,\alpha, n, n{-}1) &\stackrel{\rm HEFT}{\longrightarrow}A(1 \alpha, -v) \, = \, -A(1 \alpha, v)\, . 
    \end{split}
\end{align}
Using this relation  we obtain the KLT double copy relation for the HEFT amplitudes
\begin{align}
\label{MfunAA}
    M(12\ldots n{-}2,v)&=\sum_{\alpha,\beta\in S_{n-3}}\mathcal{S}(1\alpha|1\beta)
    A(1\alpha,v)A(1\beta, v)\, . 
\end{align}\black
We then write  the colour-ordered HEFT amplitudes with two massive scalars and $n{-}2$ gluons as \cite{Bern:2008qj,Vaman:2010ez,Du:2011js,BjerrumBohr:2012mg,Bjerrum-Bohr:2010pnr} 
\begin{align}\label{AfunN}    
A(1\alpha,v)=
\sum_{\beta\in S_{n-3}}
\mathsf{m}(1\alpha|1\beta)\npre([1\beta], v)\, ,  
\end{align}
 where  $\npre([1\beta], v)$ are KLT numerators 
  in  the Del Duca-Dixon-Maltoni (DDM)  basis \cite{DelDuca:1999rs}. This basis is associated with fully left-nested  commutators where the first index is fixed to be 1, which we denote as
  \begin{equation}\label{birdsnest}
    [1\beta]\coloneqq[1,\beta_2,\beta_3,\ldots, \beta_{n{-}2}]\coloneqq [\ldots[[[1,\beta_{2}],\beta_{3}],\beta_{4}],\ldots ,\beta_{n{-}2}]\,,
\end{equation}
where $\beta$ is a permutation of the remaining gluon legs $2,\ldots,n{-}2$.
The matrix $\mathsf{m}$ appearing in \eqref{AfunN} is the propagator matrix 
\cite{Vaman:2010ez}, 
or inverse of the KLT matrix, satisfying%
\footnote{Note that in YM theory the propagator matrix $\mathsf{m}$ is singular, while in our case it is invertible because the sum of the gluon momenta is not lightlike. }
\begin{align}
\sum_{\beta^\prime\in S_{n-3}}\mathcal{S}(1\alpha|1\beta^\prime) \mathsf{m}(1\beta^\prime|1\beta)  = \delta_{\alpha \beta}\, . 
    \end{align}
Plugging  \eqref{AfunN} into \eqref{MfunAA} we get 
\begin{align}
\begin{split}
\label{mS}
    M(12\ldots n{-}2,v)&=\sum_{\alpha,\beta, 
    \beta^\prime\in S_{n-3}}A(1\alpha,v) \mathcal{S}(1\alpha|1\beta) 
    \mathsf{m}(1\beta|1\beta^\prime)
   \npre([1\beta^\prime], v)\\
    &=\sum_{\alpha,\alpha^\prime\in S_{n-3}}
    \mathsf{m}(1\alpha|1\alpha^\prime)\npre([1\alpha^\prime], v)
     \npre([1\alpha], v)\, .
    \end{split}
\end{align}
Now, the last line of \eqref{mS}  can be recast in the form of \eqref{eq:newDC-bis}.  This can be seen from the definition of the propagator matrix \cite{Vaman:2010ez} in the  bi-adjoint scalar theory \cite{Du:2011js,BjerrumBohr:2012mg,Cachazo:2013iea,Arkani-Hamed:2017mur,Bahjat-Abbas:2018vgo}, 
\begin{align}
    \mathcal{A}_{\text{bi-adj}} = \sum_{\Gamma\in \widetilde\rho} \frac{c_\Gamma \, \widetilde c_\Gamma}{d_\Gamma}= \sum_{\alpha, \alpha^\prime \in S_{n-2}} c([1\alpha], n) \mathsf{m} (1\alpha | 1\alpha^\prime) \widetilde c([1\alpha^\prime], n)\, ,
\end{align}
and then  using the colour-kinematics duality to replace the factors of $c$ in those formulae by factors of $\npre$, thus transforming  \eqref{mS} into \eqref{eq:newDC-bis}. Here $c$ and  $\widetilde c$ are nothing but the trace of the colour group generators for the external  scalars,  that is   
\begin{align}
    c([1\alpha], n) = f^{a_1 a_{\alpha(2)} x_1}
f^{x_1 a_{\alpha(3)} x_2}\cdots  f^{x_{n-3} a_{\alpha (n-1)} a_n}\, , 
\end{align}
with a similar formula for $\widetilde c$.

Alternatively, we use the BCJ form \eqref{eq:newDC} to rewrite the gauge theory amplitude appearing in \eqref{mS}, giving
\begin{align}\label{eq: halfwayDDM}
\begin{split}
    M(12\ldots n{-}2,v)&=\sum_{\alpha\in S_{n-3}} \sum_{\Gamma \in \rho_{\alpha}}
    \frac{\npre(\Gamma, v)}{d_{\Gamma}}
     \npre([1\alpha], v)\, .
    \end{split}
\end{align}
One can then show that    \eqref{eq: halfwayDDM} can be cast in the form of  \eqref{eq:newDC-bis}. For instance, at five points \eqref{eq: halfwayDDM} reads
\begin{align}\label{eq: gravityinBCJform}
\begin{split}
    M(123,v) &=\, \left(\frac{\npre([[1,2],3],v)}{s_{12}s_{123}}+ \frac{\npre([1,[2,3]],v)}{s_{23}s_{123}}\right)\npre([[1,2],3],v) \\&+ \left(\frac{\npre([[1,3],2],v)}{s_{1
    3}s_{123}}+ \frac{\npre([1,[3,2]],v)}{s_{23}s_{123}}\right)\npre([[1,3],2],v)\\
    &=\, \frac{\npre([[1,2],3],v)^2}{s_{12}s_{123}} + \frac{\npre([1,[2,3]],v)^2}{s_{23}s_{123}} + \frac{\npre([[1,3],2],v)^2}{s_{1
    3}s_{123}
    }\, , 
\end{split}
\end{align}
where we have used anti-symmetry in $2\leftrightarrow 3$ to combine the terms with propagator $\frac{1}{s_{23}}$ using the Jacobi identity 
\begin{align}
    \npre([[1,2],3],v)-\npre([[1,3],2],v)=\npre([1,[2,3]],v)\, .
\end{align}
The final line of \eqref{eq: gravityinBCJform} is precisely in the form of  \eqref{eq:newDC-bis}.  In general, at higher points, we can always use Jacobi identities to combine terms in this manner.

For completeness we also   derive the  propagator matrix $\mathsf{m} (1\alpha | 1\alpha^\prime)$ in the five-point example.  From \eqref{eq:newDC} we have 
\begin{align}
\begin{split}
\label{5ptex}
    A(123,v) &= \frac{\npre([[1,2],3],v)}{s_{12}s_{123}}+\frac{\npre([1,[2,3]],v)}{s_{23}s_{123}}\ , \\
    A(132,v) &= \frac{\npre([[1,3],2],v)}{s_{13}s_{123}}+\frac{\npre([1,[3,2]],v)}{s_{23}s_{123}} \ .
\end{split}    
\end{align}
Not all the numerators above are in  the DDM basis (or fully left-nested), and thus we  rewrite them using the kinematic Jacobi identities as
\begin{align}
\begin{split}
\npre([1,[2,3]],v) &= \npre([1,2,3],v) - \npre([1,3,2],v)\ , \\ 
\npre([1,[3,2]],v) &= -\npre([1,2,3],v) + \npre([1,3,2],v) \, .
 \end{split} 
\end{align}
Plugging these into \eqref{5ptex} we have
\begin{align}
\begin{pmatrix}A(123,v)\\ A(132,v)
\end{pmatrix} = \mathsf{m}(1\{23\}|1\{23\})
\begin{pmatrix}\npre([1,2,3],v)\\ \npre([1,3,2],v)
\end{pmatrix}
\ , 
\end{align}
where 
\begin{align}
\mathsf{m}(1\{23\}|1\{23\})=    \frac{1}{s_{123}}
\begin{pmatrix}
\frac{1}{s_{12}}+ \frac{1}{s_{23}}  &  -\frac{1}{s_{23}}
\\ -\frac{1}{s_{23}} & \frac{1}{s_{13}}+ \frac{1}{s_{23}}
\end{pmatrix}\, .
\end{align}
It is immediate to check that  $\mathsf{m}(1\{23\}|1\{23\})
$ 
is non-singular since $s_{123} {=} (p_1{+}p_2{+}p_3)^2{\neq}0  $, as mentioned earlier.

\subsection{Comparison to Yang-Mills theory}

It is interesting to compare \eqref{AfunN} to the corresponding formula for an $n$-point gluon amplitude in Yang-Mills theory 
\begin{align}
\label{eq: AfunNYM}  
{\cal A}(1,\alpha,n)=
\sum_{\beta\in S_{n-2}}
\mathsf{m}(1\alpha|1\beta)\npre([1,\beta], n)\, .  
    \end{align}
Here the expansion is over the $(n{-}2)!$ fully left-nested DDM numerators $\npre([1,\beta], n)$ with particles $1$ and $n$ fixed, which are widely studied in the literature \cite{Stieberger:2016lng,Nandan:2016pya,delaCruz:2016gnm,Schlotterer:2016cxa,Fu:2017uzt,Teng:2017tbo,Du:2017kpo}. Each term in the sum corresponds to a half-ladder, or multi-peripheral graph, and  in this case the matrix $\mathsf{m}(1\alpha|1\beta)$ is singular. 
When we take the decoupling limit \eqref{dec-limit}, 
mapping our HEFT amplitudes (with two scalars and $n{-}2$ gluons) into $(n{-}1)$-point YM  amplitudes, the  YM amplitudes thus obtained  are in the form \eqref{eq: AfunNYM} (with $n$ replaced by $n{-}1$).

We also note the alternative KLT representation of YM amplitudes with three legs being fixed \cite{Bjerrum-Bohr:2010pnr,Kiermaiertalk}, 
\begin{align}
\label{eq: AfunNYM-bis}  
{\cal A}(1,2, \alpha,n)=
\sum_{\beta\in S_{n-3}}
\mathsf{m}(12\alpha|12\beta)\widetilde{\npre}([1,2,\beta], n)\, ,  
    \end{align}
where the matrix $\mathsf{m}(12\alpha|12\beta)$ is now invertible, which implies that the $\widetilde{\npre}$ are gauge invariant and unique, but may contain poles \cite{Chen:2017bug}. The $\widetilde \npre$ represent a basis of numerators which are equivalent to the $\npre$ up to the kernel of the non-invertible matrix $\mathsf{m}(1\alpha|1\beta)$ with $\alpha,\beta \in S_{n-2}$. However, we have used generalised gauge transformations to explicitly set linearly dependent numerators to zero as per \cite{Kiermaiertalk}.

As an example, we review the four-point case. Here we have, from 
\eqref{5ptex}, 
\begin{align}
\begin{pmatrix}{\cal A}(1,2,3,4)\\ {\cal A}(1,3,2,4)
\end{pmatrix} &=
\begin{pmatrix}
\frac{1}{s_{12}}+ \frac{1}{s_{23}}  &  -\frac{1}{s_{23}}
\\ -\frac{1}{s_{23}} & \frac{1}{s_{13}}+ \frac{1}{s_{23}}
\end{pmatrix}\begin{pmatrix}\npre([1,2,3],4)\\ \npre([1,3,2],4)
\end{pmatrix}\, \nn\\
&=\begin{pmatrix}
\frac{1}{s_{12}}+ \frac{1}{s_{23}}  &  -\frac{1}{s_{23}}
\\ -\frac{1}{s_{23}} & \frac{1}{s_{13}}+ \frac{1}{s_{23}}
\end{pmatrix}\Bigg[\begin{pmatrix}\npre([1,2,3],4)\\ \npre([1,3,2],4)
\end{pmatrix}
+\begin{pmatrix}{s_{12}\over s_{13}}\npre([1,3,2],4)\\ -\npre([1,3,2],4)
\end{pmatrix}\Bigg]\nn\\
&=\begin{pmatrix}
\frac{1}{s_{12}}+ \frac{1}{s_{23}}  &  -\frac{1}{s_{23}}
\\ -\frac{1}{s_{23}} & \frac{1}{s_{13}}+ \frac{1}{s_{23}}
\end{pmatrix}\begin{pmatrix}\widetilde\npre([1,2,3],4)
\\ 0
\end{pmatrix}\,  , 
\end{align}
so that $\widetilde \npre([1,2,3], 4)\!=\!\npre([1,2,3], 4)+{s_{12}\over s_{13}}\npre([1,3,2], 4)$,  and $\widetilde \npre([1,3,2], 4)\!=\!0$. 
Note that, in contrast to the numerators $\npre$, the numerators $\widetilde{\npre}$ are not
crossing symmetric.

\section{Pre-numerators and quasi-shuffle products}
\label{sec:3}
The BCJ numerators introduced in the previous section can be generated from an object known as the pre-numerator \cite{Chen:2019ywi, Chen:2021chy, Brandhuber:2021bsf}, denoted in the same way as the BCJ numerators but without the commutator structure:
\begin{equation}\label{eq: prenumcanonicalordering}
    \cN(123\ldots n\!-\!2,v)\,.
\end{equation}
We can construct the BCJ numerators by permuting the labels of the pre-numerator according to the commutator structure, for example 
\begin{equation}\label{eq: bcjfromprenum}
   \begin{tikzpicture}[baseline={([yshift=-0.8ex]current bounding box.center)}]\tikzstyle{every node}=[font=\small]    
   \begin{feynman}
    \vertex (a)[myblob]{};
     \vertex [above=0.3cm of a](b)[dot]{};
     \vertex [left=0.6cm of b](c);
     \vertex [left=0.22cm of b](c23);
     \vertex [above=0.14cm of c23](v23)[dot]{};
    \vertex [above=.4cm of c](j1){$1$};
    \vertex [right=.7cm of j1](j2){$2$};
    \vertex [right=0.5cm of j2](j3){$3$};
   	 \diagram*{(a) -- [thick] (b),(b) -- [thick] (j1),(v23) -- [thick] (j2),(b)--[thick](j3)};
    \end{feynman}  
  \end{tikzpicture}\!\!:\quad  \cN([[1,2], 3],v):= \cN(123,v)-\cN(213,v)+\cN(321,v)-\cN(312,v)\,,
\end{equation}
and similarly for more complicated BCJ numerators. Written in this way the BCJ numerators manifestly satisfy the Jacobi identities.

 The BCJ numerators are unique, however the pre-numerators used to build them are not uniquely defined: there is still freedom in choosing their  exact form, which we can use   to make certain properties of the pre-numerators manifest. Remarkably, it was shown  in \cite{Brandhuber:2021bsf} that for tree amplitudes in HEFT there exists a closed-form expression for \emph{crossing symmetric} pre-numerators in terms of 
gauge-invariant quantities for arbitrary~$n$. Moreover, 
  the pre-numerators  in the canonical gluon ordering $\npre(12\ldots n{-}2,v)$ can be  constructed from  a quasi-shuffle algebra using  a two-step procedure which we now outline, following~\cite{Brandhuber:2021bsf}.%
\footnote{One can  then define the pre-numerators with other orderings via the simple relabelling
      $
      \npre(12\ldots j\ldots i\ldots n{-}2, v):= \npre(12\ldots i\ldots j\ldots n{-}2, v)|_{i\leftrightarrow j}
 $. 
This definition also has the benefit of making crossing symmetry manifest.}
\black

First,  we build an intermediate object which we call an ``algebraic'' pre-numerator \cite{Chen:2019ywi,Chen:2021chy,Brandhuber:2021eyq,Brandhuber:2021bsf}, 
\begin{align}
    \widehat\npre(123\ldots n{-}2):=T_{(1)}\star T_{(2)}\star T_{(3)}\star \cdots \star T_{(n{-}2)}\,, 
\end{align}
obtained by fusing abstract generators $T_{(i)}$, one for each gluon leg, using a quasi-shuffle product denoted by $\star$. Then we define a linear map (denoted by angle brackets $\langle \bullet \rangle$) from these abstract generators to kinematic quantities
\begin{equation}\label{eq: anglebracketmap}
    \npre(123\ldots n{-}2, v):=\big \langle \widehat\npre(123\ldots n{-}2)\big \rangle\,.
\end{equation}
The linear map relates each single-index generator, 
e.g.~$\langle T_{(i)}\rangle $, to a vector current with a heavy source. Multi-index generators are   mapped to  multi-rank tensor currents of the same heavy source. The fusion product of these currents can be understood as an algebraic fusion rule from lower-rank to higher-rank currents once we perform the map \eqref{eq: anglebracketmap}. In this sense, the algebra generators are identified as the operators that generate the tensor currents \cite{Chen:2019ywi,Chen:2021chy}.   

In the next two sections we review  the quasi-shuffle product, before describing the map $\langle\bullet\rangle$ to tensor currents and finally pre-numerators.

\subsection{The quasi-shuffle  product}
\label{sec:3.1}
The quasi-shuffle algebra consists of a vector space of generators $\mathfrak{A}$, with generators written as
\begin{equation}
    T_{(\tau_1),\ldots(\tau_r)}\, , 
\end{equation}
labelled by subsets of the gluon indices $\tau_i\subset \{1,2,\ldots ,n{-}2\}$, for example $T_{(12),(45),(3)}$. As sets, the $\tau_i$ are unordered and therefore the labels contained \textit{within} each $\tau_i$ can be written in any order by definition;  we choose for convenience to write them in the order $\{1,2,3,\ldots\}$. However, the subsets themselves $(\tau_1),\ldots,(\tau_r)$ are ordered.

To begin, we introduce some standard nomenclature for generators: we will refer to generators with a single subset $T_{(\tau_1)}$ as \emph{``letters''}, those with  multiple subsets $T_{(\tau_1),\ldots,(\tau_r)}$ as \emph{``words''},  and  the length of a word
is the number of subsets, or letters, it contains.

Next we introduce the quasi-shuffle product $\star$ on the space of generators $\mathfrak{A}$, first in some examples and then in full generality. 
The simplest case to consider is the product of two letters, say, $T_{(1)}$ and $T_{(2)}$, which gives the four-point algebraic pre-numerator:
\begin{align}\label{eq: shuffleproduct}
    T_{(1)}\star T_{(2)} = T_{(1),(2)}+T_{(2),(1)}-T_{(12)}\,.
\end{align}
The first two terms above correspond to the ``shuffle'' part of the quasi-shuffle product, while the last term is a letter made from ``stuffing'' $(1)$ and $(2)$ together (hence the {\it quasi}-shuffle nature of the algebra). The ``stuffing'' terms can be understood as giving rise to  ``contact term'' corrections to the lower-point rules such that the numerators constructed in this way lead to the correct amplitudes. For instance, at four points, naively attaching two three-point numerators together by shuffling would lead to $T_{(1),(2)}$ and $T_{(2),(1)}$, and the required correction  to obtain the correct numerator (or equivalently amplitude) is precisely $T_{(12)}$.%
\footnote{We are grateful to Oliver Schlotterer for a discussion on this point.} 

The product of a letter with a word is also fairly straightforward and follows the same splitting into shuffle and stuffing terms.  As an  example, consider
 \begin{align}\label{eq: shuffleproduct2}
 \begin{split}
     T_{(1)}\star T_{(2),(3),(4)} &= T_{(1),(2),(3),(4)} + T_{(2),(1),(3),(4)}+ T_{(2),(3),(1),(4)}+ T_{(2),(3),(4),(1)} \\
     &- T_{(12),(3),(4)} -T_{(2),(13),(4)}-T_{(2),(3),(14)}\,.
 \end{split}
 \end{align}
 Note that this product  preserves the ordering of the letters in the word,  $T_{(2),(3),(4)}$. This is true in general: the quasi-shuffle of two words $T_{(\tau_1),(\tau_2)\ldots (\tau_r)}$ and  $T_{(\rho_1),(\rho_2)\ldots (\rho_r)}$  preserves the ordering of the $\tau_i$ and the $\rho_j$. The general  formula for the product of two arbitrary words is given by \cite{Brandhuber:2021bsf}
 \begin{align}\label{eq: stuffleExplicit}
    T_{(\tau_1), \ldots ,(\tau_r)}\star T_{(\rho_1), \ldots ,(\rho_s)} = \sum_{ \genfrac{}{}{0pt}{}{\sigma\lvert_{\{\tau\}} =\{(\tau_1), \ldots ,(\tau_r)\}}{\sigma\lvert_{\{\rho\}} =\{(\rho_1), \ldots ,(\rho_s)\} } }
    (-1)^{t-r-s} T_{(\sigma_1), \ldots ,(\sigma_t)}\, , 
\end{align}
where the $\tau_i$ or $\rho_i$ are now any subsets of $\{1,\ldots,n{-}2\}$. The notation $\sigma\lvert_{\{\tau\}}$ means that we restrict the partition $\sigma$ onto the subset $\tau \!=\! \tau_1\cup\tau_2\cup \cdots \cup \tau_r $, for example $\{(234),(56),(78)\} \lvert_{\{2,4,6\}} = \{(24),(6)\}$. There is also a recursive definition of the quasi-shuffle product \cite{hoffman2000quasi, hoffman2017quasi} which we give for completeness in 
Appendix~\ref{app: RecursiveDefs}. The quasi-shuffle product defined in this way is associative and commutative, hence we can perform the products in any order we choose.

\subsection{Mapping to the pre-numerator}\label{sec: MapToKinematics}

After constructing the algebraic pre-numerator $\widehat \npre(12\ldots n{-}2)$ using \eqref{eq: stuffleExplicit}, we then use the angle-bracket map to obtain  the pre-numerator (from which we finally obtain BCJ numerators as e.g.~in \eqref{eq: bcjfromprenum}). The key step is clearly the $\langle \bullet \rangle$ map, which we now discuss.  

The angle-bracket map for the special case of a single label generator is simply 
\begin{align}
    \langle T_{(i)}\rangle =v\Cdot \eps_i \, , 
\end{align}
while  for a generic   generator, it is given by \cite{Brandhuber:2021bsf}
\begin{align}\label{eq: extendedBracket2}
         \langle T_{(\tau_1),(\tau_2),\ldots,(\tau_r)} \rangle = \frac{v\Cdot F_{\tau_1}\Cdot V_{\Theta(\tau_2)}\Cdot F_{\tau_2}\ldots \Cdot V_{\Theta(\tau_r)}\Cdot F_{\tau_r}\Cdot v}{(n-2)v\Cdot p_{\tau_1 [1]} v\cdot p_{\tau_1}\ldots v\cdot p_{\tau_{r-1}}}\,, 
 \end{align} 
 where   the length of $ \tau_1\cup\cdots\cup\tau_{r}$ is $n{-}2$.   
 $\Theta(\tau_i)$ is the subset of labels in $ \tau_1\cup\cdots\cup\tau_{i-1}$ which are smaller than the first label in $\tau_i$ for the canonical ordering $1,2,\ldots, n{-}2$, and $\tau_1[1]$ is the first label in $\tau_1$.
As an example, if $T_{(13),(25),(46)}$ then $\Theta(25)=\{1\}, \Theta(46)=\{1,2,3\}$ which can also be seen by locating all the labels to the south-west of the labels $2$ and $4$ in the following ``musical diagram''  \cite{Brandhuber:2021bsf}
 \begin{align}
     \begin{tikzpicture}[baseline={([yshift=-0.8ex]current bounding box.center)}]\tikzstyle{every node}=[font=\small]    
   \begin{feynman}
    \vertex (l1)[]{}; \vertex [left=0.5cm of l1](rm1)[]{$(\tau_1)$};\vertex [right=4.cm of l1](r1)[]{}; \vertex [right=0.5cm of l1](v1)[sb]{\white\textbf\small $\mathbf{1}$};\vertex [right=1.5cm of l1](v4)[sb]{\white\textbf\small $\mathbf{3}$};
    \vertex [above=0.5cm of l1](l2)[]{};  \vertex [right=4.0cm of l2](r2)[]{};\vertex [left=0.5cm of l2](rm2)[]{$(\tau_2)$};\vertex [right=1.cm of l2](v2)[sb]{\white\textbf\small $\mathbf{2}$};\vertex [right=2.5cm of l2](v6)[sb]{\white\textbf\small $\mathbf{5}$};
     \vertex [above=0.5cm of l2](l3)[]{};  \vertex [right=4.0cm of l3](r3)[]{};\vertex [left=0.5cm of l3](rm3)[]{$(\tau_3)$};\vertex [right=2.0cm of l3](v3)[sb]{\white\textbf\small $\mathbf{4}$};\vertex [right=3.0cm of l3](v7)[sb]{\white\textbf\small $\mathbf{6}$};
   	 \diagram*{(l1)--[thick](v1)- -[thick](v4)--[thick](r1),(l2)--[thick](v2)--[thick](v6)--[thick](r2),(l3)--[thick](v3)--[thick](v7)--[thick](r3)};
    \end{feynman}  
  \end{tikzpicture}
 \end{align}
Here we have also defined $V_\tau^{\mu\nu}= v^\mu p_{\tau}^\nu$,  and $F_{\tau_i}$ is the product of linearised field strengths $F_j$ for all $j \in \tau_i$ where $\tau_j$ is ordered with respect to the canonical ordering $1,2,\ldots, n{-}2$. For example, $F_{123}=F_1\Cdot F_2 \Cdot F_3$.  
Finally, if any of the sets $\Theta(\tau_i)$ happens to be empty, then we set that generator to zero. 
 
 Note that here we always consider the canonical gluon ordering $1,2,\ldots, n{-}2$ when we take the map $\langle\bullet \rangle$. This is what makes the algebraic pre-numerator $\widehat \npre(12\ldots n{-}2)$ symmetric in the gluon labels,  while  the kinematic pre-numerator $ \npre(12\ldots n{-}2)$  is not. In Section~\ref{sec:non-abelian} we will consider an extension to the algebra which includes arbitrary gluon orderings, however for now  we will use the canonical ordering.
 
For the first few lower-point cases, the pre-numerators are given explicitly by
\begin{align}\label{eq: prenumExamples}
    \npre(1,v)&= \langle T_{(1)} \rangle = v\Cdot\varepsilon_1\, , \\
    \begin{split}
    \label{n12}
    \npre(12,v)&= \langle T_{(1)}\star T_{(2)}\rangle =\cancelto{0}{\langle T_{(1),(2)}\rangle} + \cancelto{0}{\langle T_{(2),(1)}\rangle} - \langle T_{(12)}\rangle \\ & =  -\frac{v\Cdot F_{1}\Cdot F_{2}\Cdot v}{2 v \Cdot p_1}\, , 
    \end{split}\\
    \begin{split}\label{eq: prenumExamples-2}
    \npre(123,v)&= \langle\cancelto{0}{T_{\text{(1),(2),(3)}}}\rangle- \langle\cancelto{0}{T_{\text{(1),(23)}}}\rangle-\langle T_{\text{(12),(3)}}\rangle+ \langle\cancelto{0}{T_{\text{(1),(3),(2)}}}\rangle\\ 
    & 
    -\langle T_{
   \text{(13),(2)}}\rangle+ \langle\cancelto{0}{T_{\text{(2),(1),(3)}}}\rangle- \langle\cancelto{0}{T_{\text{(2),(13)}}}\rangle+\langle\cancelto{0}{T_{\text{(2),(3),(1)}}}\rangle\\
   & - \langle\cancelto{0}{T_{
   \text{(23),(1)}}}\rangle+ \langle\cancelto{0}{T_{\text{(3),(1),(2)}}}\rangle- \langle\cancelto{0}{T_{\text{(3),(12)}}}\rangle+ \langle\cancelto{0}{T_{\text{(3),(2),(1)}}}\rangle+\langle T_{
   \text{(123)}}\rangle\\
   & = -\frac{v\Cdot F_1 \Cdot F_2 \Cdot V_{12}\Cdot F_{3}\Cdot v}{3 v\Cdot p_1 v\Cdot p_{12}} -  \frac{v\Cdot F_1 \Cdot F_3 \Cdot V_{1}\Cdot F_{2}\Cdot v}{3 v\Cdot p_1 v\Cdot p_{13}} + \frac{v\Cdot F_1 \Cdot F_2 \Cdot F_3\Cdot v}{3 v\Cdot p_1}\,.
\end{split}
\end{align}
In the above expressions, we have made  explicit the terms that are mapped to zero according to the rules \eqref{eq: extendedBracket2}. 
From these examples it seems as though there is a large   redundancy  in the pre-numerator, since  many terms are set to zero by the map. Despite this redundancy, we shall see in the next section that these terms are essential for linking the coproduct to the factorisation behaviour of the pre-numerator.%
\footnote{There is an alternative construction where we treat gluon $1$  as special, and in that construction the vanishing terms do not appear \cite{Brandhuber:2021bsf}.}

\section{Kinematic Hopf algebra for the pre-numerator}\label{sec:Hopf-algebra}

The quasi-shuffle product described in the last section gives us an algorithmic way to produce  $n$-point pre-numerators and hence all BCJ numerators in the HEFT. The quasi-shuffle algebra can be extended to a bialgebra with the introduction of two new operations: a coproduct: $\Delta$, and a counit: $\epsilon$, which must satisfy specific compatibility conditions we will detail later in this section. Additionally, there exists another operation, the antipode: $S$, which further extends this bialgebra to a Hopf algebra, and is also subject to various compatibility conditions. In this section we will describe each one of these new operations in turn, starting with the coproduct and its relation to
(multiple) factorisation of the pre-numerators.

\subsection{The coproduct and factorisation}

The coproduct is a linear map from the space  $\mathfrak{A}$ of generators $T_{\omega}$, introduced in Section~\ref{sec:3.1},   to the tensor product space $\mathfrak{A}\otimes \mathfrak{A}$. For the quasi-shuffle product, the coproduct is well known \cite{hoffman2017quasi,hoffman2000quasi} and can be derived from two key relations. First, we define the coproduct of the simplest object, a letter
\begin{equation}\label{eq: cop1}
    \Delta(T_{(\tau)}) = \Id\otimes T_{(\tau)} + T_{(\tau)}\otimes \Id\,,
\end{equation}
where $\Id$ is the identity element of $\star$, the ``empty'' word. Then we extend the definition to products of letters using 
a defining property of the coproduct: compatibility with the quasi-shuffle product,
\begin{equation}\label{eq: cop2}
    \Delta(T_{(\tau_1)\ldots(\tau_r)}\star T_{(\rho_1)\ldots(\rho_s)}) = \Delta(T_{(\tau_1)\ldots(\tau_r)})\star \Delta(T_{(\rho_1)\ldots(\rho_s)})\,,
\end{equation}
where the extension of the product to tensors is defined naturally as $(A\otimes B)\star (C\otimes D)=(A\star C)\otimes(B\star D)$. We can represent the above equation (and similar equations later on) diagrammatically as follows:
\begin{align}\label{eq: diagramcoproductcompatible}
\begin{tikzpicture}[baseline={([yshift=-1.5ex]current bounding box.center)}]\tikzstyle{every node}=[font=\small]    
   \begin{feynman}
    \vertex (i1){};
    \vertex [right=0.8cm of i1](ic){};
    \vertex [right=1.6cm of i1](i2){};
    \vertex [below=.9cm of ic](v1)[dot]{};
    \vertex [below=1.6cm of v1](v2)[codot]{};
    \vertex [below=3.2cm of i1](o1);
    \vertex [below=3.2cm of i2](o2);
   	 \diagram*{(i1) -- [thick, bend right] (v1),(i2) -- [thick, bend left] (v1),(v2) -- [thick, bend right] (o1),(v2) -- [thick,bend left] (o2),(v1)--[thick](v2)};
    \end{feynman}  
  \end{tikzpicture}
  ~~~~=~~~~ 
     \begin{tikzpicture}[baseline={([yshift=-1.5ex]current bounding box.center)}]\tikzstyle{every node}=[font=\small]    
   \begin{feynman}
    \vertex (i1){};
    \vertex [right=1.6cm of i1](i2){};
    \vertex [below=.8cm of i1](v1)[codot]{};
    \vertex [below=.8cm of i2](v12)[codot]{};
    \vertex [below=.8cm of v1](vc);
    \vertex [below=.8cm of v12](vc2);
    \vertex [below=.8cm of vc](v2)[dot]{};
    \vertex [below=.8cm of vc2](v22)[dot]{};
    \vertex [below=.8cm of v2](o1);
    \vertex [below=.8cm of v22](o2);
    \vertex [left=.7cm of vc](vm1);
    \vertex [right=.8cm of vc](vm2);
    \vertex [right=.7cm of vc2](vm3);
    \vertex [left=0.65cm of vc2](vm4){};
   	 \diagram*{(i1) -- [thick] (v1),(v2) -- [thick] (o1),(i2) -- [thick] (v12),(v22) -- [thick] (o2),(v1) -- [thick,bend right] (vm1),(vm1) -- [thick,bend right] (v2),(v1) -- [thick,bend left] (vm2),(vm2) -- [thick,bend right] (v22),(v12) -- [thick,bend right] (vm4),(vm4) -- [thick,bend left] (v2),(v12) -- [thick,bend left] (vm3),(vm3) -- [thick,bend left] (v22)};
    \end{feynman}  
  \end{tikzpicture}\, ,
\end{align}
where the red dot $\textcolor{red}{\bullet}$ denotes the coproduct and the black dot ${\bullet}$  the quasi-shuffle product. Note that such diagrams should be read from top to bottom.

Using \eqref{eq: cop1} and \eqref{eq: cop2} we can deduce the coproduct of the algebraic pre-numerator~as 
\begin{equation}\label{eq: coproductPrenum}
    \Delta(T_{(1)}\star T_{(2)}\star \cdots \star T_{(n{-}2)}) = \Delta(T_{(1)})\star \Delta(T_{(2)})\star\cdots\star \Delta(T_{(n-2)})\,.
\end{equation}
Additionally, we can consider the coproduct of an arbitrary word (which will in general not be a product of other words)
\begin{align}\label{eq: coproductDef}
    \cop(T_{(\tau_1),\ldots, (\tau_r )}):=\sum_{i=0}^{r}T_{(\tau_1),\ldots, (\tau_i )}\otimes T_{(\tau_{i+1}),\ldots, (\tau_r )}\, .
\end{align}
Equation \eqref{eq: coproductPrenum} is  usually the fastest way to compute the coproduct, but first it is worth using the explicit definition \eqref{eq: coproductDef} in an example. Consider the coproduct of the four-point algebraic pre-numerator,
\begin{equation}\label{eq: 4pCoprod}
\begin{split}
     \Delta(\widehat \cN(12))&= 
     \Delta(T_{(1)}\star T_{(2)})= 
     \Delta(T_{(1)(2)}) + \Delta(T_{(2)(1)})-\Delta(T_{(12)})\\
     &=\Id\otimes T_{(1)(2)} + T_{(1)}\otimes T_{(2)} +  T_{(1)(2)}\otimes \Id \\& 
     +\Id \otimes  T_{(2)(1)} + T_{(2)}\otimes T_{(1)} +\Id\otimes  T_{(2)(1)}\\
     &-\Id\otimes T_{(12)} - T_{(12)}\otimes \Id\,.
\end{split}
\end{equation}
We can also introduce $\Delta'$, the reduced coproduct, which simply removes all {\it trivial} terms involving  the identity $\Id$. For the four-point pre-numerator this leaves us with
\begin{align}
     \Delta'(\widehat \cN(12)) =  &T_{(1)}\otimes T_{(2)} +  T_{(2)}\otimes T_{(1)} =\widehat \cN(1)\otimes \widehat \cN(2) + \widehat \cN(2)\otimes \widehat \cN(1)
     \, .
\end{align}
 In this example there are two important points to emphasise:
 \begin{itemize}
     \item[{\bf 1.~}] 
     The coproduct has the general property of splitting up words (and algebraic pre-numerators) which, we will see, is reminiscent of factorisation on the massive propagators $v\Cdot p_{\tau}$. 
     \item[{\bf 2.}] 
     The non-trivial terms in the coproduct comes precisely  from  the terms $T_{(1),(2)}$ and $T_{(2),(1)}$ which did not contribute to the pre-numerator upon taking the map $\langle \bullet\rangle$ in  \eqref{eq: prenumExamples}. 
     \end{itemize}
This is true in general: we can derive an expression for the coproduct of the $n$-point algebraic pre-numerator in terms of tensor products of lower-point ones, 
\begin{align}\label{eq: coprodPrenum}
    \cop\widehat\npre(123\ldots n{-}2)&=\sum_{\sigma_L\cup\sigma_R=\{1,\ldots,n{-}2\}}\widehat\npre(\sigma_L)\otimes \widehat\npre(\sigma_R)\, ,
\end{align}
where here we allow $\sigma_L$ and $\sigma_R$ to be empty and define $\widehat\npre(\emptyset):= \Id$. Thus if we demand $\sigma_L$, $\sigma_R$ to be non-empty in the above equation we get $\cop'(\widehat\npre(123\ldots n{-}2))$ instead. The proof of the above formula comes directly from the relation \eqref{eq: coproductPrenum}. Indeed,  assuming  that the $(n{-}1)$-point pre-numerator takes the form above, then by induction for the $n$-point pre-numerator we have   
\begin{align}
\begin{split}
  \cop \widehat\npre(123\ldots n{-}2)&= \cop(\widehat\npre(123\ldots n{-}3)\star T_{(n-2)})=\cop(\widehat\npre(123\ldots n{-}3))\star\cop(T_{(n-2)})\\
&  =\sum_{\sigma_L\cup\sigma_R=\{1,2,\ldots,n{-}3\}}(\widehat\npre(\sigma_L)\otimes \widehat\npre(\sigma_R))\star (\Id\otimes T_{(n-2)}+T_{(n-2)}\otimes\Id)\\
&=\sum_{\sigma_L\cup\sigma_R=\{1,2,\ldots,n{-}2\}}(\widehat\npre(\sigma_L)\otimes \widehat\npre(\sigma_R)) \, .
\end{split}
\end{align}
To make the connection between the coproduct and factorisation precise we need to extend the map $\langle\bullet\rangle$ to include tensor structures. To do this let us first consider the factorisation property of the four-point pre-numerator 
\begin{equation}\label{eq :4pFactor}
    \cN(12,v)= -\frac{v\Cdot F_1\Cdot F_2 \Cdot v}{2 v\Cdot p_1}\,.
\end{equation}
The factorisation  of the above pre-numerator as  $v\Cdot p_1 \!\rightarrow\! 0$ is characterised by the corresponding residue: 
\begin{equation}
\begin{split}
    \Res_{(v\Cdot p_1)}\cN(12,v) &= v\Cdot \varepsilon_1 \frac{p_1\Cdot p_2}{2} v \Cdot \varepsilon_2 =\cN(1,v)\frac{p_1\Cdot p_2}{2}  \cN(2,v)\,.
\end{split}
\end{equation}
 More generally, the pre-numerator has the following factorisation behaviour, proven in \cite{Brandhuber:2021bsf},  obtained from  the residues at the poles $v\Cdot p_{1\tau}{=} 0$:
\begin{equation}\label{eq: prenumFactor}
    \Res_{(v\Cdot p_{1\tau})} \cN(12\ldots n{-}2,v) =\frac{\|\tau \| \|\omega\|}{n-2} \cN(\tau,v) (p_{\Theta(\omega)}\Cdot p_{\omega[1]})  \cN(\omega,v)\,,
\end{equation}
where $\tau$ and $\omega$ are subsets of $\{2,3,\ldots, n{-}2\}$ such that $\tau\cup \omega = \{2,3,\ldots,{n-2}\}$, and $\| \tau \|$, $\| \omega \|$ are their respective lengths. 
Again,  $\Theta(\omega)$ is the set of labels in $\{1,2,\ldots, n{-}2\}$ which are less that $\omega[1]$.   Note that in the HEFT due to momentum conservation we have $v\Cdot p_{1\ldots n{-}2}=0$; for example at five points we could write $v\Cdot p_{12}= - v\Cdot p_3$. Thus, in the above we have included leg $1$ in $v\Cdot p_{1\tau}$ to fix such ambiguities.

Now we come to a crucial point:   each one of these massive poles can be matched to a term in the coproduct if we  define a replacement rule $\mathcal{C}$  for tensor products as  
\begin{align}\label{eq: repRuleC}
   \mathcal{C}(T_{\omega_1}\otimes T_{\omega_2})&=
   \frac{\|\omega_1 \| \|\omega_2\|}{\|\omega_1 \|+\|\omega_2\|}
   \pole{v\Cdot p_{\omega_1}}
     T_{\omega_1} p_{\Theta(\omega_2)}\Cdot p_{\omega_2[1]}  T_{\omega_2}\,, 
\end{align}
where $\Theta$ is defined in the same way as before, and if $\Theta(\omega)=\emptyset$ then we set $p_\emptyset=0$. Note that the algebraic generators have not yet been evaluated, and should be treated as commuting objects.
Finally, to map to physical quantities we must apply the angle bracket $\langle\bullet\rangle$ to the generators, as in 
\begin{equation}
    \langle T_{\omega_1}T_{\omega_2}\rangle =  \langle T_{\omega_1}\rangle\langle T_{\omega_2}\rangle\, .
\end{equation}
To make the connection between factorisation and the coproduct, we must take the residue at the relevant poles before we perform the angle bracket map to kinematics. Explicitly, we have the following relation for any subset $\tau\subset \{2,\ldots,n-2\}$:
\begin{align}\label{eq: copAlgPre}
\Big\langle \Res_{(v\Cdot p_{1\tau})} \mathcal{C} \cop' \widehat\npre(12\ldots n{-}2) \Big\rangle=  \Res_{(v\Cdot p_{1\tau})} \npre(12\ldots n{-}2,v) 
   \, . 
\end{align}
We also note that the pre-numerator has poles which are not colour ordered e.g.~when $v \Cdot p_{13}=0$, and therefore the residues at these poles must cancel when we combine these pre-numerators into 
colour-ordered amplitudes.

In summary, the coproduct decomposes a pre-numerator into a sum of terms, each of which corresponds to a factorisation channel on a single massive pole. It is useful to illustrate the connection between coproducts and factorisation in some examples. 

$\bm{n{=}4}$. In this case,  we calculated the coproduct of the algebraic pre-numerator in  \eqref{eq: 4pCoprod}.  Mapping this result to physical quantities using \eqref{eq: repRuleC} we obtain
 \begin{align}
 \begin{split}
   \cC\cop'\widehat\npre(12)&= \cC(T_{{(1)}}\otimes T_{{(2)}}) + \cancelto{0}{\cC(  T_{{(2)}}\otimes T_{{(1)}}}) \\
     &=\frac{1}{2}\pole{v\Cdot p_1} T_{(1)}  p_1\Cdot p_2  T_{(2)}\, .
     \end{split}
 \end{align}%
 Taking the residue at the pole $v\Cdot p_1= 0$, and mapping  to physical quantities  we get 
 \begin{equation}
       \langle\Res_{v\Cdot p_1}\cC\cop'\widehat\npre(12)\rangle =\frac{1}{2}\langle T_{(1)}  p_1\Cdot p_2  T_{(2)} \rangle = \frac{1}{2}\langle T_{(1)}\rangle  p_1\Cdot p_2  \langle T_{(2)} \rangle =\frac{1}{2} v\Cdot\varepsilon_1 p_1\Cdot p_2 v\Cdot \varepsilon_2\, , 
 \end{equation}
 where the right-hand side is nothing but the factorisation behaviour of the four-point pre-numerator in \eqref{eq :4pFactor}. 
 
$\bm{n{=}5}$. At  five points  the algebraic pre-numerator is given in  \eqref{eq: prenumExamples-2}, and its coproduct is 
\begin{align}
\begin{split}
  \cop\widehat\npre(123)
  &=\Id\otimes \widehat\npre(123) + \widehat\npre(123)\otimes \Id \\
  &+ \widehat\npre(1)\otimes \widehat\npre(23)+ \widehat\npre(2)\otimes \widehat\npre(13)+ \widehat\npre(3)\otimes \widehat\npre(12)\\
  &+ \widehat\npre(23)\otimes \widehat\npre(1)+ \widehat\npre(13)\otimes \widehat\npre(2)+ \widehat\npre(12)\otimes \widehat\npre(3)\, .
\end{split}
\end{align}
 Proceeding as before, we act  on the above equation with the replacement rules $\cC$ as given in \eqref{eq: repRuleC}, to  find that only $\widehat\npre(1)\otimes \widehat\npre(23)$, $\widehat\npre(13)\otimes \widehat\npre(2)$, and $\widehat\npre(12)\otimes \widehat\npre(3)$ are non-vanishing. Thus we get
\begin{align}
\begin{split}
     \cC\cop'(\widehat\npre(123))&=\cC\Big( \widehat\npre(1)\otimes \widehat\npre(23)+ \widehat\npre(13)\otimes \widehat\npre(2)+ \widehat\npre(12)\otimes \widehat\npre(3)\Big) \\
    &=\frac{2}{3}\pole{v\Cdot p_1}\widehat\npre(1) p_1\Cdot p_2 \widehat\npre(23)+\frac{2}{3}\pole{v\Cdot p_{13}}\widehat\npre(13) p_1\Cdot p_2 \widehat\npre(2) \\
    &+\frac{2}{3}\pole{v\Cdot p_{12}}\widehat\npre(12) p_{12}\Cdot p_3 \widehat\npre(3)\,.
    \end{split}
\end{align}
Now if we take the residue at  the pole, say, $v\cdot p_{13}= 0$, and apply the map $\langle\bullet\rangle$ we obtain
\begin{equation}
    \langle\Res_{v\Cdot p_{13}}\,\mathcal{C}\cop' \widehat\npre(123) \rangle=  {v\Cdot F_1\Cdot F_3\Cdot v\over 3 \,v\Cdot p_1} p_1\Cdot p_2  v\Cdot \varepsilon_2\,,
\end{equation}
which is the expected factorisation of the pre-numerator in the limit $v\Cdot p_{13}\to 0$.

We now move on to discuss iterated coproducts and their connection to generalised factorisation. 
\subsection{Generalised factorisation and iterated coproducts}\label{sec: Itercoproduct}

We now  examine the factorisation of the pre-numerators when we take multiple massive propagators on shell -- we call this generalised factorisation.   Interestingly, the result will depend on the order in which we take the limits, in contrast to amplitudes, where the order of limits is irrelevant. As an example consider the five-point pre-numerator
\begin{equation}
    \begin{split}
    \npre(123,v)&= -\frac{v\Cdot F_1 \Cdot F_2 \Cdot V_{12}\Cdot F_{3}\Cdot v}{3 v\Cdot p_1 v\Cdot p_{12}} -  \frac{v\Cdot F_1 \Cdot F_3 \Cdot V_{1}\Cdot F_{2}\Cdot v}{3 v\Cdot p_1 v\Cdot p_{13}} + \frac{v\Cdot F_1 \Cdot F_2 \Cdot F_3\Cdot v}{3 v\Cdot p_1}\,.
\end{split}
\end{equation}
There are three massive poles we can consider here which we write as $v\Cdot p_1\!\rightarrow\!0$, $v\Cdot p_{12}\!\rightarrow\!0$ and $v\Cdot p_{13}\!\rightarrow\!0$ (recall that $v\Cdot p_{123}=0$ here so that $v\Cdot p_2= -v\Cdot p_{13}$). If we take $v\Cdot p_1\rightarrow 0$ we find, using  \eqref{eq: prenumFactor}
\begin{equation}
    \Res_{(v\Cdot p_1)}\cN(123,v)=\frac{2}{3} \npre(1,v)p_{1}\Cdot p_{2}\npre(23,v)\, .
\end{equation}
 Then taking $v\Cdot p_{2}\rightarrow0$, which is now equivalent to $v\Cdot p_{12}\rightarrow 0$ we obtain 
 \begin{equation}\label{eq: multiPoleEx}
     \Res_{(v \Cdot p_{2},\, v\Cdot p_1)}\cN(123,v)=\frac{1}{3}\npre(1,v)p_1\Cdot p_2 \cN(2,v) p_2\Cdot p_3 \cN(3,v)\,.
 \end{equation}
 Note  that the labels appearing in $\Res_{(v \Cdot p_{2},\, v\Cdot p_1)}$ are ordered. Alternatively, we could first take the limit $v\Cdot p_{12}\rightarrow 0$ then $v\Cdot p_1 \rightarrow 0$ which first gives
\begin{equation}
    \Res_{(v\Cdot p_{12})}\cN(123,v)= \frac{2}{3}\npre(12,v)p_{12}\Cdot p_{3}\npre(3,v)\,,
\end{equation}
 and then 
 \begin{equation}
          \Res_{(v \Cdot p_{1},\, v\Cdot p_{12})}\cN(123,v)=\frac{1}{3}\npre(1,v)p_1\Cdot p_2\cN(2,v) p_{12}\Cdot p_3 \cN(3,v)\,.
 \end{equation}
 Comparing the above with  \eqref{eq: multiPoleEx} we see that although the pre-numerator has been split into the same pre-numerators ($\cN(1,v)$, $\cN(2,v)$ $\cN(3,v)$), the momentum products differ. All of the different choices of multiple factorisations of $\npre(123,v)$ can be summarised in the following diagram:
  \begin{equation*}
   \hspace*{-0.6cm}
    \begin{tikzcd}[cramped, column sep=0.6em]
        & &  \npre(123,v)\arrow[dl, "v\Cdot p_{1}\rightarrow 0"'] \arrow[d, "v\Cdot p_{12}\rightarrow 0"] \arrow[dr,"v\Cdot p_{13}\rightarrow 0"] &   \\
        & \frac{2}{3}\npre(1,v)p_{1}\Cdot p_{2} \npre(23,v) \arrow[d,"v\Cdot p_2\rightarrow 0"] & \frac{2}{3} \npre(12,v)p_{12}\Cdot p_{3} \npre(3,v)  \arrow[d,"v\Cdot p_1\rightarrow 0"] & \frac{2}{3} \npre(13,v)p_{1}\Cdot p_{2} \npre(2,v)  \arrow[d,"v\Cdot p_1\rightarrow 0"]\\
        & \frac{1}{3}p_{1}\Cdot p_{2}p_{2}\Cdot p_{3}\prod_{i=1}^3 \npre(i,v) &   \frac{1}{3}p_{1}\Cdot p_{2}p_{12}\Cdot p_{3}\prod_{i=1}^3 \npre(i,v) & \frac{1}{3} p_{1}\Cdot p_{3} p_{1}\Cdot p_{2}\prod_{i=1}^3 \npre(i,v)
    \end{tikzcd}
    \\
 \end{equation*}
 This process continues for higher-point pre-numerators,  and in general the order in which the limits are taken is important. By applying the coproduct to a pre-numerator several times  we aim to build an object which can recreate this factorisation structure by taking multiple residues, as we now discuss.

If we apply $k$ coproducts to a generator we land in the 
 $(k+1)$-tensor product space of $\mathfrak{A}$, which we write as  $\mathfrak{A}^{\otimes (k+1)}$. 
 A $k$-iterated coproduct is denoted as $\Delta^{(k)}$ and defined recursively as
\begin{align}\label{eq: iterCop}
    \cop^{(k)}=(\cop^{(k-1)}\otimes \Id)\cop\,.
\end{align} 
 For example $\cop^{(2)} \, T_{(1),(2),(3)}$ is given by
\begin{align}
\begin{split}
    \cop^{(2)}T_{(1),(2),(3)}&=(\cop\otimes \Id )\cop(T_{(1),(2),(3)})\\
    &=(\cop\otimes \Id )\Big[\Id\otimes T_{(1),(2),(3)} + T_{(1)}\otimes T_{(2)(3)}+ T_{(1),(2)}\otimes T_{(3)} +  T_{(1)(2)(3)}\otimes \Id\Big]\\
    &= \Id \otimes \Id \otimes T_{(1),(2),(3)} + \Id \otimes T_{(1)}\otimes T_{(2),(3)}+ T_{(1)} \otimes \Id\otimes T_{(2),(3)}\\
    & + \Id \otimes T_{(1),(2)}\otimes T_{(3)}+ T_{(1)}\otimes T_{(2)}\otimes T_{(3)}+ T_{(1),(2)}\otimes \Id\otimes T_{(3)}\\
    &+\Id \otimes T_{(1),(2),(3)}\otimes \Id +T_{(1)} \otimes T_{(2),(3)}\otimes \Id +T_{(1),(2)} \otimes T_{(3)}\otimes \Id
    \\
    &+T_{(1),(2),(3)} \otimes \Id \otimes \Id\,,
    \end{split}
\end{align}
where in the second equality we have used $\cop(\Id)=\Id\otimes\Id$. As before,  we can introduce  a reduced iterated  coproduct $\Delta'^{(k)}$, which simply removes the trivial terms involving  the identity. For the example above this gives 
\begin{equation}
    \cop'^{(2)}T_{(1),(2),(3)}=T_{(1)} \otimes T_{(2)}\otimes T_{(3)}\,.
\end{equation}
Note that the choice of $(\cop\otimes \Id)$ instead of $(\Id\otimes \cop)$ in  \eqref{eq: iterCop} is arbitrary, due to the co-associativity of the coproduct:
\begin{equation}\label{eq: coassociativity}
     (\Id\otimes \cop) \cop (T_{\omega})=(\cop \otimes \Id)\cop (T_{\omega}) \,.
\end{equation}
 This property can also be illustrated diagrammatically as
\begin{align}\label{eq: diagramcoassociativity}
\begin{tikzpicture}[baseline={([yshift=-1.5ex]current bounding box.center)}]\tikzstyle{every node}=[font=\small]    
   \begin{feynman}
    \vertex (i1){};
    \vertex [below=0.8cm of i1](c1)[codot]{};
    \vertex [below left=1.3cm of c1](v1);
    \vertex [below right=1.3cm of c1](c2)[codot]{};
    \vertex [below  =0.95cm of v1](o1){};
    \vertex [below left =1.3cm of c2](o2);
    \vertex [below right=1.3cm of c2](o3);
   	 \diagram*{(i1) -- [thick] (c1),(c1) -- [thick, bend right] (v1)[] -- [thick] (o1),(c1) -- [thick, bend left] (c2),(c2)--[thick, bend right](o2),(c2)--[thick, bend left](o3)};
    \end{feynman}  
  \end{tikzpicture}
  ~~~=~~~
  \begin{tikzpicture}[baseline={([yshift=-1.5ex]current bounding box.center)}]\tikzstyle{every node}=[font=\small]    
   \begin{feynman}
    \vertex (i1){};
    \vertex [below=0.8cm of i1](c1)[codot]{};
    \vertex [below right=1.3cm of c1](v1);
    \vertex [below left=1.3cm of c1](c2)[codot]{};
    \vertex [below  =0.95cm of v1](o1){};
    \vertex [below right =1.3cm of c2](o2);
    \vertex [below left=1.3cm of c2](o3);
   	 \diagram*{(i1) -- [thick] (c1),(c1) -- [thick, bend left] (v1)[] -- [thick] (o1),(c1) -- [thick, bend right] (c2),(c2)--[thick, bend left](o2),(c2)--[thick, bend right](o3)};
    \end{feynman}  
  \end{tikzpicture}
  \,.
\end{align}
Using the general form of a single coproduct of the algebraic pre-numerators \eqref{eq: coprodPrenum}, and the recursive definition for the iterated coproduct $\cop^{(k)}$ in \eqref{eq: iterCop}, we can deduce the action of $\Delta^{(k)}$ on a generic algebraic pre-numerator: 
\begin{align}
  \cop^{(k)}(\widehat\npre(12\ldots n{-}2))= \sum_{\tau_1\cup\cdots \cup\tau_k\cup\tau_{k+1}= \{1,\ldots,n-2\}} \widehat\npre(\tau_1)\otimes \widehat\npre(\tau_2)\otimes\cdots\otimes \widehat\npre(\tau_{k+1})
   \, , 
\end{align}
where  again we allow any of the $\tau_i$ to be empty, and we can obtain the expression for $\cop'^{(k)}$ by demanding that the $\tau_i$ be non-empty.

We can now naturally extend the replacement rule $\cC$ defined in \eqref{eq: repRuleC} to include an arbitrary number of tensor products,
   \begin{align}
   \begin{split}
       \mathcal{C}\Big(  T_{\omega_1}\otimes T_{\omega_2}\otimes \cdots\otimes T_{\omega_i}\Big)& :=\frac{\|\omega_1 \| \|\omega_2\|\ldots \|\omega_i\|}{\|\omega_1 \|+\|\omega_2\|+\ldots\|\omega_i\|}\pole{v\Cdot p_{\omega_1}v\Cdot p_{\omega_1\omega_2}\ldots v\Cdot p_{\omega_1\cdots \omega_{i-1}}}\\
       &\times\Big( T_{\omega_1} p_{\Theta(\omega_2)}\Cdot p_{\omega_2(1)} T_{\omega_2} \cdots p_{\Theta(\omega_i)}\Cdot p_{\omega_i(1)} T_{\omega_i}\Big)\, ,
       \end{split}
   \end{align}
where $\Theta(\omega_i)$ is the set of elements in $\omega_1\cup\cdots\cup \omega_{i-1}$ which are less than the first element of $\omega_{i}$ in the canonical ordering $(1,2\ldots, n{-}2)$. Now we can explicitly obtain $k$ factorisations of a pre-numerator from  the $k$ coproduct by taking the residues on the appropriate poles and applying the map $\langle\bullet\rangle$, 
\begin{align}\label{eq: kcopAlgPre}
 \Big\langle \Res_{(v\Cdot p_{\omega_1},\ldots v\Cdot p_{\omega_k} )} \cC\cop'^{(k)}\widehat\npre(12\ldots n{-}2)\Big\rangle=  \Res_{(v\Cdot p_{\omega_1},\ldots v\Cdot p_{\omega_k} )}\Big(\npre(12\ldots n{-}2,v)\Big)
   \, ,
\end{align}
where once again both of these residues are ordered and taken right to left.

As a nontrivial example, consider the five-point case. After applying the first coproduct one obtains 
\begin{align}
\begin{split}
 &\cop^{(2)}(\widehat\npre(123))= (\cop\otimes \Id) \cop(\widehat\npre(123))\\
 & =  \cop(\Id)\otimes \widehat\npre(123) + \cop(\widehat\npre(123))\otimes \Id \\
  &+ \cop(T_{(1)})\otimes \widehat\npre(23)+ \cop(T_{(2)})\otimes \widehat\npre(13)+ \cop(T_{(3)})\otimes \widehat\npre(12)\\
  &+ \cop(\widehat\npre(23))\otimes T_{(1)}+ \cop(\widehat\npre(13))\otimes T_{(2)}+ \cop(\widehat\npre(12))\otimes T_{(3)}\,,
\end{split}
\end{align}
from which we can identify which pieces contribute to factorisation by using the replacement rule $\mathcal{C}$ defined in \eqref{eq: repRuleC}. Doing so, we see that the terms which contribute~are
\begin{align}
\begin{split}
     \mathcal{C} \cop'^{(2)}(\widehat\npre(123)) &=\mathcal{C}\Big(\cop(\widehat\npre(12))\otimes \widehat\npre(3)\Big)  + \mathcal{C}\Big(\cop(\widehat\npre(13))\otimes \widehat\npre(2)\Big)  \\
    &= \mathcal{C}\Big(T_{(1)}\otimes T_{(2)}\otimes T_{(3)}\Big) + \mathcal{C}\Big(T_{(1)}\otimes T_{(3)}\otimes T_{(2)} \Big)\\
    &=\frac{1}{3} \pole{v\Cdot p_1 v\Cdot p_{12}}T_{(1)}p_{1}\Cdot p_2 T_{(2)} p_{12}\Cdot p_{3} T_{(3)} \\&+\frac{1}{3}\pole{v\Cdot p_1 v\Cdot p_{13}}T_{(1)}p_{1}\Cdot p_{3}T_{(3)} p_{1}\Cdot p_{2}T_{(2)}\, .
\end{split}
\end{align}
Extracting the residue corresponding to  $v\Cdot p_1\rightarrow 0 $ and then  $v\Cdot p_2\rightarrow 0$, we get 
\begin{align}
\begin{split}
  & \Res_{(v\Cdot p_2,v\Cdot p_1)}\cC \cop'^{(2)}\widehat\npre(123)\\
   &=\oint_{v\Cdot p_2=0} {d(v\Cdot p_2)}\oint_{v\Cdot p_1=0} {d(v\Cdot p_1)}\frac{1}{3}\bigg(\frac{T_{(1)}p_{1}\Cdot p_2 T_{(2)} p_{12}\Cdot p_{3} T_{(3)}}{v\Cdot p_1 v\Cdot p_{12}}  +\frac{T_{(1)}p_{1}\Cdot p_{3}T_{(3)} p_{1}\Cdot p_{2}T_{(2)}}{v\Cdot p_1 v\Cdot p_{13}}\bigg)\\
   &=\frac{1}{3}\, T_{(1)} \, p_1\Cdot p_2 \, T_{(2)}   \, p_{2}\Cdot p_3 \,  T_{(3)}\,,
   \end{split}
\end{align}
on which we can act with the map $\langle \bullet \rangle$ to obtain the factorisation 
\begin{align}
\begin{split}
    \big\langle \Res_{(v\Cdot p_2,v\Cdot p_1)}\cC \cop'^{(2)}\widehat\npre(123,v) \big\rangle &=\frac{1}{3} \, v\Cdot \varepsilon_1 \, p_1\Cdot p_2 \,v\Cdot \varepsilon_2   \, p_{2}\Cdot p_3 \,  v\Cdot \varepsilon_3\,\\
    &=\Res_{(v\Cdot p_2,v\Cdot p_1)}\,\npre(123,v) \, .
    \end{split}
\end{align}
The other two factorisations can be  obtained similarly: 
\begin{align}
\begin{split}
    \big\langle\Res_{(v\Cdot p_1,v\Cdot p_{12})} \cC\cop'^{(2)}\widehat\npre(123)&\big\rangle =\frac{1}{3}v\Cdot \varepsilon_1 \, p_1\Cdot p_2 \, v\Cdot \varepsilon_2  \, p_{12}\Cdot p_3 \, v\Cdot \varepsilon_3\, ,  \\
    \big\langle\Res_{(v\Cdot p_1,v\Cdot p_{2})} \cC\cop'^{(2)}\widehat\npre(123)\big\rangle&=-\frac{1}{3}v\Cdot \varepsilon_1 \, p_1\Cdot p_3 \, v\Cdot \varepsilon_2 \,  p_{1}\Cdot p_2 \, v\Cdot \varepsilon_3\, .
    \end{split}
\end{align}
It is easy to check that the three residues are not linearly independent:
\begin{align}
\begin{split}
  &\langle \Res_{(v\Cdot p_2,v\Cdot p_1)} \cC \cop'^{(2)}(\widehat\npre(123))\rangle\\
  =& \, \langle \Res_{(v\Cdot p_1,v\Cdot p_{12})} \cC \cop'^{(2)}(\widehat\npre(123))\rangle+ \langle \Res_{(v\Cdot p_1,v\Cdot p_{2})} \cC \cop'^{(2)}(\widehat\npre(123))\rangle\, ,
\end{split}
\end{align}
which is a consequence of the global residue theorem of multi-dimensional integrals~\cite{griffiths2014principles}.

As a final comment, we note that for an $n$-point amplitude we can have non-trivial iterated coproducts $\cop^{(k)}$ with  $k=2, \ldots, n{-}3$. In  the context of multiple polylogarithms,  $\cop^{(n-3)}$ corresponds to what is usually known as the symbol \cite{Goncharov:2010jf}. Note however that in the case of iterated integrals, the relevant algebra is a shuffle algebra \cite{ree-shuffle}.%
\footnote{Interestingly, quasi-shuffle algebras also make an appearance in the context of multiple polylogarithms when  written in terms of nested sums rather than iterated integrals, see Section~3.4 of \cite{Duhr:2014woa} for a review. We also discuss the connection between quasi-shuffle and shuffle algebras in  Appendix \ref{app:B}.}

\subsection{The counit}
To obtain the quasi-shuffle bialgebra we  introduce the counit $\epsilon$, defined by its action on the generators $T_{\omega}$ and the identity $\Id$:
\begin{align}\label{eq: counitDef}
    \epsilon(\Id):=\Id,\qquad  \epsilon(T_{\omega}):=0\,.
\end{align}
The counit can be used to ``undo'' the action of the coproduct as follows:
\begin{equation}\label{eq: counitRel1}
    \star( \Id\otimes\epsilon) \cop(T_{\omega})=\star(\epsilon\otimes \Id) \cop(T_{\omega})=T_{\omega}\,,
\end{equation}
where by $\star$ we mean $\star(T_{\omega_1}\otimes T_{\omega_2}) = T_{\omega_1}\star T_{\omega_2}$. Again, the content of this equation can be shown using diagrams:
\begin{align}\label{eq: diagramcounitRel1}
    \begin{tikzpicture}[baseline={([yshift=-1.5ex]current bounding box.center)}]\tikzstyle{every node}=[font=\small]    
   \begin{feynman}
    \vertex (in){};
    \vertex [below=.8cm of in](v1)[codot]{};
    \vertex [below=.8cm of v1](vc);
    \vertex [right=0.4cm of vc](v2)[sb]{\white\textbf\small $\mathbf{\epsilon}$};
    \vertex [left=0.6cm of vc](v3);
    \vertex [below=.8cm of vc](v4)[dot]{};
    \vertex [below=.8cm of v4](v5){};
   	 \diagram*{(in) -- [thick] (v1),(v1) -- [thick,bend left] (v2),(v1) -- [thick,bend right] (v3),(v2) -- [thick,bend left] (v4),(v3) -- [thick,bend right] (v4), (v4) --[thick](v5)};
    \end{feynman}  
  \end{tikzpicture}~~~=~~~\begin{tikzpicture}[baseline={([yshift=-1.5ex]current bounding box.center)}]\tikzstyle{every node}=[font=\small]    
   \begin{feynman}
    \vertex (in){};
    \vertex [below=.8cm of in](v1)[codot]{};
    \vertex [below=.8cm of v1](vc);
    \vertex [left=0.4cm of vc](v2)[sb]{\white\textbf\small $\mathbf{\epsilon}$};
    \vertex [right=0.6cm of vc](v3);
    \vertex [below=.8cm of vc](v4)[dot]{};
    \vertex [below=.8cm of v4](v5){};
   	 \diagram*{(in) -- [thick] (v1),(v1) -- [thick,bend right] (v2),(v1) -- [thick,bend left] (v3),(v2) -- [thick,bend right] (v4),(v3) -- [thick,bend left] (v4), (v4) --[thick](v5)};
    \end{feynman}  
  \end{tikzpicture}~~~=~~~\begin{tikzpicture}[baseline={([yshift=-1.5ex]current bounding box.center)}]\tikzstyle{every node}=[font=\small]    
   \begin{feynman}
    \vertex (i1){};
    \vertex [below=3.2cm of i1](v5){};
     \vertex [above=1.2cm of i1](v6){};
   	 \diagram*{(i1) -- [thick] (v5)};
    \end{feynman}  
  \end{tikzpicture}\,.
\end{align}
One can also check directly that the counit is compatible with the quasi- shuffle product,
\begin{align}\label{eq: counitRel2}
    \epsilon(T_{\omega_1}\star T_{\omega_2})=\epsilon(T_{\omega_1})\star \epsilon(T_{\omega_2})\,,
\end{align}
which can also be summarised in a diagram
\begin{align}\label{eq: diagramcounitRel2}
    \begin{tikzpicture}[baseline={([yshift=-1.5ex]current bounding box.center)}]\tikzstyle{every node}=[font=\small]    
   \begin{feynman}
    \vertex (i1){};
    \vertex [right=0.6cm of i1](ic){};
    \vertex [right=1.2cm of i1](i2){};
    \vertex [below=.9cm of ic](v1)[dot]{};
    \vertex [below=1.cm of v1](v2)[sb]{\white\textbf\small $\mathbf{\epsilon}$};
    \vertex [below=0.9cm of v2](o1);
   	 \diagram*{(i1) -- [thick, bend right] (v1),(i2) -- [thick, bend left] (v1),(v1)--[thick](v2), (v2)--[thick](o1)};
    \end{feynman}  
  \end{tikzpicture}~~~=~~~\begin{tikzpicture}[baseline={([yshift=-1.5ex]current bounding box.center)}]\tikzstyle{every node}=[font=\small]    
   \begin{feynman}
    \vertex (i1){};
    \vertex [right=0.6cm of i1](ic){};
    \vertex [right=1.2cm of i1](i2){};
    \vertex [below=.6cm of ic](v1u);
    \vertex [below=0.9cm of i1](v2)[sb]{\white\textbf\small $\mathbf{\epsilon}$};
    \vertex [below=0.9cm of i2](v3)[sb]{\white\textbf\small $\mathbf{\epsilon}$};
    \vertex [below=1.2cm of v1](v4)[dot]{};
    \vertex[below=1.0cm of v4](v5)[]{};
   	 \diagram*{(i1) -- [thick] (v2),(i2) -- [thick] (v3),(v3) -- [thick,bend left] (v4),(v2) -- [thick,bend right] (v4),(v4)--[thick](v5)};
    \end{feynman}  
  \end{tikzpicture}\,\,.
 \end{align}
Unlike the coproduct, the counit does not have a particularly useful interpretation in terms of a property of the pre-numerator, since it simply maps the algebraic pre-numerator to zero,
\begin{equation}
    \epsilon(\widehat\npre(12\ldots n{-}2,v))=0\,,
\end{equation}
which is easily seen from the definition \eqref{eq: counitDef}.

\subsection{The antipode}

The final operation we need to introduce to form a Hopf algebra is the antipode $S$,  which is a linear map from the space of generators $\mathfrak{A}$ to itself. Its defining property is \begin{align}\label{antipodedefproperty}
    \star(\Id\otimes S)\cop(T_{\omega})=\star(S\otimes\Id)\cop(T_{\omega})=\epsilon(T_{\omega})\,,
\end{align}
which can also be described diagrammatically as
\begin{align}\label{eq: diagramantipodedefproperty}
    \begin{tikzpicture}[baseline={([yshift=-1.5ex]current bounding box.center)}]\tikzstyle{every node}=[font=\small]    
   \begin{feynman}
    \vertex (in){};
    \vertex [below=.8cm of in](v1)[codot]{};
    \vertex [below=.8cm of v1](vc);
    \vertex [right=0.4cm of vc](v2)[sb]{\white\textbf\small $\mathbf{s}$};
    \vertex [left=0.6cm of vc](v3);
    \vertex [below=.8cm of vc](v4)[dot]{};
    \vertex [below=.8cm of v4](v5){};
   	 \diagram*{(in) -- [thick] (v1),(v1) -- [thick,bend left] (v2),(v1) -- [thick,bend right] (v3),(v2) -- [thick,bend left] (v4),(v3) -- [thick,bend right] (v4), (v4) --[thick](v5)};
    \end{feynman}  
  \end{tikzpicture}~~~=~~~\begin{tikzpicture}[baseline={([yshift=-1.5ex]current bounding box.center)}]\tikzstyle{every node}=[font=\small]    
   \begin{feynman}
    \vertex (in){};
    \vertex [below=.8cm of in](v1)[codot]{};
    \vertex [below=.8cm of v1](vc);
    \vertex [left=0.4cm of vc](v2)[sb]{\white\textbf\small $\mathbf{s}$};
    \vertex [right=0.6cm of vc](v3);
    \vertex [below=.8cm of vc](v4)[dot]{};
    \vertex [below=.8cm of v4](v5){};
   	 \diagram*{(in) -- [thick] (v1),(v1) -- [thick,bend right] (v2),(v1) -- [thick,bend left] (v3),(v2) -- [thick,bend right] (v4),(v3) -- [thick,bend left] (v4), (v4) --[thick](v5)};
    \end{feynman}  
  \end{tikzpicture}~~~=~~~\begin{tikzpicture}[baseline={([yshift=-1.5ex]current bounding box.center)}]\tikzstyle{every node}=[font=\small]    
   \begin{feynman}
    \vertex (i1){};
     \vertex [below=1.6cm of i1](v1)[sb]{\white\textbf\small $\mathbf{\epsilon}$};
    \vertex [below=3.2cm of i1](v5){};
   	 \diagram*{(i1) -- [thick] (v1)-- [thick] (v5)};
    \end{feynman}  
  \end{tikzpicture}\,.
\end{align}
In simple terms, 
we can take a generator and first split it up via the coproduct, then act on one of the tensor product factors with $S$, and then recombine with the product. The result of this operation  is that all generators except for the identity are mapped to zero as per \eqref{eq: counitDef}.%
\footnote{The antipode of a generator gets its name by analogy with antipodal points on a sphere whose vectors sum to zero.}

The explicit action of the antipode on a generator is defined  recursively,
\begin{align}\label{eq: antipodeDef}
\begin{split}
S(T_{(\tau_1),\ldots, (\tau_r)})&:=-\sum_{i=0}^{r-1}S(T_{(\tau_1),\ldots, (\tau_i)})\star T_{(\tau_{i+1}),\ldots,(\tau_r)}\,,\\
S(\Id)&:=\Id\,.
\end{split}
\end{align}
Using this, one can check the following required compatibility properties:
\begin{align}\label{eq: antipodeRelations}
\begin{split}
    S(T_{\omega_1}\star T_{\omega_2})&= S(T_{\omega_2})\star S(T_{\omega_1})\, , \\
   ( S\otimes S)  \cop(T_{\omega})&= R_{12} \cop  S(T_{\omega})\, , \\
   \epsilon  S(T_{\omega})&=\epsilon(T_{\omega})\, ,
\end{split}
\end{align}
where $R_{12}$ denotes the exchange of  the first and second element in the tensor product. The diagrammatic representation of the above relations are as follows,
\begin{align}\label{eq: diagramAntipodeRelations}
    \begin{tikzpicture}[baseline={([yshift=-1.5ex]current bounding box.center)}]\tikzstyle{every node}=[font=\small]    
   \begin{feynman}
    \vertex (i1){};
    \vertex [right=0.6cm of i1](ic){};
    \vertex [right=1.2cm of i1](i2){};
    \vertex [below=.9cm of ic](v1)[dot]{};
    \vertex [below=1.cm of v1](v2)[sb]{\white\textbf\small $\mathbf{s}$};
    \vertex [below=1.cm of v2](o1);
   	 \diagram*{(i1) -- [thick, bend right] (v1),(i2) -- [thick, bend left] (v1),(v1)--[thick](v2), (v2)--[thick](o1)};
    \end{feynman}  
  \end{tikzpicture}~~~=~~~\begin{tikzpicture}[baseline={([yshift=-1.5ex]current bounding box.center)}]\tikzstyle{every node}=[font=\small]    
   \begin{feynman}
    \vertex (i1){};
    \vertex [right=0.6cm of i1](ic){};
    \vertex [right=1.2cm of i1](i2){};
    \vertex [below=.6cm of ic](v1){};
    \vertex [below=.6cm of ic](v1u);
    \vertex [below=1.3cm of i1](v2)[sb]{\white\textbf\small $\mathbf{s}$};
    \vertex [below=1.3cm of i2](v3)[sb]{\white\textbf\small $\mathbf{s}$};
    \vertex [below=1.5cm of v1](v4)[dot]{};
    \vertex[below=1.0cm of v4](v5)[]{};
   	 \diagram*{(i1) -- [thick,bend right](v1)-- [thick,bend left] (v3),(i2) -- [thick,bend left](v1u)-- [thick,bend right] (v2),(v3) -- [thick,bend left] (v4),(v2) -- [thick,bend right] (v4),(v4)--[thick](v5)};
    \end{feynman}  
  \end{tikzpicture}\,\,, && \begin{tikzpicture}[baseline={([yshift=-1.5ex]current bounding box.center)}]\tikzstyle{every node}=[font=\small]    
   \begin{feynman}
    \vertex (i1){};
    \vertex [below=0.9cm of i1](v0)[codot]{};
    \vertex [below=0.9cm of v0](vc1);
     \vertex [left=0.4cm of vc1](v1)[sb]{\white\textbf\small $\mathbf{s}$};
      \vertex [right=0.4cm of vc1](v2)[sb]{\white\textbf\small $\mathbf{s}$};
      \vertex [below=0.9cm of v1](o1);
      \vertex [below=0.9cm of v2](o2);
   	 \diagram*{(i1) -- [thick](v0),(v0) -- [thick, bend right](v1)--[thick](o1),(v0) -- [thick, bend left](v2)--[thick](o2)};
    \end{feynman}  
  \end{tikzpicture}~~~=~~~\begin{tikzpicture}[baseline={([yshift=-1.5ex]current bounding box.center)}]\tikzstyle{every node}=[font=\small]    
   \begin{feynman}
    \vertex (i1){};
    \vertex [below=0.6cm of i1](vm1)[sb]{\white\textbf\small $\mathbf{s}$};
    \vertex [below=0.6cm of vm1](v0)[codot]{};
    \vertex [below=0.5cm of v0](vc1);
     \vertex [left=0.5cm of vc1](v1);
      \vertex [right=0.5cm of vc1](v2);
       \vertex [below=0.65cm of vc1](vc2);
       \vertex [below=0.5cm of vc1](vc2u){};
       \vertex [below=1.2cm of v1](o1);
       \vertex [below=1.2cm of v2](o2);
   	 \diagram*{(i1) -- [thick](vm1) -- [thick](v0),(v0)--[thick,bend right](v1),(v0) -- [thick,bend left](v2),(v1) -- [thick,bend right](vc2)-- [thick,bend left](o2),(v2) -- [thick,bend left](vc2u)-- [thick,bend right](o1)};
    \end{feynman}  
  \end{tikzpicture}\,\,,&&
  \begin{tikzpicture}[baseline={([yshift=-1.5ex]current bounding box.center)}]\tikzstyle{every node}=[font=\small]    
   \begin{feynman}
    \vertex (i1){};
    \vertex [below=1.0cm of i1](v0)[sb]{\white\textbf\small $\mathbf{s}$};
     \vertex [below=1.2cm of v0](v1)[sb]{\white\textbf\small $\mathbf{\epsilon}$};
    \vertex [below=3.2cm of i1](v5){};
   	 \diagram*{(i1) -- [thick] (v0)-- [thick] (v1)-- [thick] (v5)};
    \end{feynman}  
  \end{tikzpicture}~~~=~~~ \begin{tikzpicture}[baseline={([yshift=-1.5ex]current bounding box.center)}]\tikzstyle{every node}=[font=\small]    
   \begin{feynman}
    \vertex (i1){};
     \vertex [below=1.6cm of i1](v1)[sb]{\white\textbf\small $\mathbf{\epsilon}$};
    \vertex [below=3.2cm of i1](v5){};
   	 \diagram*{(i1) -- [thick] (v1)-- [thick] (v5)};
    \end{feynman}  
  \end{tikzpicture}\,\,.
\end{align}
To find the action of the antipode on the algebraic pre-numerator we first use \eqref{eq: antipodeDef} to find the antipode of a single letter,
\begin{equation}
    S(T_{(i)}) = -T_{(i)}\,,
\end{equation}
then we use the first relation of \eqref{eq: antipodeRelations} to find
\begin{equation}
\begin{split}
    S \widehat{\cN}(12\ldots n{-}2) &=  S(T_{(1)}\star T_{(2)}\star T_{(3)}\cdots T_{(n{-}2)}) \\
    &= S(T_{(n-2)})\star S(T_{(n-3)})\star S(T_{(n-4)})\star \cdots \star S(T_{(1)}) \\
    &= (-1)^{n-2} T_{(1)}\star T_{(2)}\star T_{(3)}\star \cdots\star T_{(n{-}2)}\\
    &= (-1)^{n-2}\widehat{\cN}(12\ldots n{-}2)\, ,
\end{split}
\end{equation}
where in the second-to-last line we have use the fact that the quasi-shuffle product is commutative. Thus the antipode only changes the pre-numerator by at most an overall sign. This seems to suggest that, like the counit, there is no useful interpretation of the antipode. However, we shall see in the next section that we can extend the quasi-shuffle algebra 
to a non-abelian version in which case the antipode 
becomes non-trivial.

\section{The non-abelian quasi-shuffle Hopf algebra} \label{sec:non-abelian}

In Section~\ref{sec:3} we laid out properties of the quasi-shuffle product used to build pre-numerators and, in turn, BCJ numerators. 
As an explicit example consider the four-point BCJ numerator
\begin{equation}
    \npre([1,2],v) =  \npre(12,v)-\npre(21,v)\,.
\end{equation}
$\npre(12,v)$ was given in \eqref{n12}, 
and to find $\npre(21,v)$ we simply swap the labels $1$ and $2$ of the pre-numerator $\npre(12,v)$. Thus the four-point  BCJ numerator is given by
\begin{equation}
    \npre([1,2],v)= -\frac{v\Cdot F_{1}\Cdot F_2 \Cdot v}{2v\Cdot p_1}+\frac{v\Cdot F_{2}\Cdot F_1 \Cdot v}{2v\Cdot p_2}= -\frac{v\Cdot F_{1}\Cdot F_2 \Cdot v}{v\Cdot p_1}\, , 
\end{equation}
where in the last equality we have used $(p_{1}+p_{2})\Cdot v  =0$. 

Note that in this example  the pre-numerator $\npre(12,v)$ is not symmetric in the labels $1$ and $2$, whereas the {\it algebraic} pre-numerator $\widehat\npre(12)= T_{(1)}\star T_{(2)}$ is symmetric since the quasi-shuffle product is commutative. The ordering of labels $(1,2)$ is only imposed on the pre-numerator once we take the angle bracket map $\langle\bullet \rangle$ given in \eqref{eq: extendedBracket2}. This raises the interesting question of whether we can incorporate arbitrary orderings of legs, beyond the canonical ordering $1,2,\ldots, n{-}2$, into the algebra itself while preserving the Hopf algebra structure.

As we now demonstrate it is possible to extend the 
quasi-shuffle product to become non-abelian by endowing our generators with additional dependence on the overall ordering of the massless particles denoted by $\alpha \in S_{n-2}$. 
First steps in this direction were taken in the appendix of the prequel paper \cite{Brandhuber:2021bsf}, but here we
will present the complete construction, the non-abelian kinematic Hopf algebra, and find, in particular, an amplitude interpretation for its coproduct and the antipode. The important result will be the ability to compute BCJ numerators directly at the level of the non-abelian quasi-shuffle algebra. 

\subsection{Extended generators}
\label{sec:5.1}

The first important step is to introduce
{\it extended} generators with ordering $\alpha \in S_{n-2}$ of the form\footnote{ Remarkably, an extension of the generators analogous to this one has led to the kinematic algebra construction for the BCJ numerators of the scattering amplitudes as well as form factors in Yang-Mills-scalar theory coupled with a bi-adjoint $\phi^3$ interaction with a general mass~\cite{Chen:2022nei}.}
\begin{align}\label{eq: extendedGen}
    T^{(\alpha)}_{(\tau_1),(\tau_2)\ldots(\tau_r)} := r^{(\alpha)} \otimes  T_{(\tau_1),(\tau_2)\ldots(\tau_r)}\,,
\end{align}
where the $\tau_i$ are subsets of $\{1,2,\ldots, n{-}2\}$. In practice the actual labels appearing in the subscript $\tau=\tau_1\cup \cdots \cup \tau_r$ will be a subset of $\alpha$, for example $T^{(12435)}_{(12),(34)}$.
These new generators  correspond to a tensor product extension of the underlying vector space of the quasi-shuffle algebra. As implied by \eqref{eq: extendedGen}, we extend the generators involving indices $\{1,2,\ldots, n{-}2\}$ by tensoring with the free algebra over the elements of $\{1,2,\ldots, n{-}2\}$. This algebra is generated by taking products and linear combinations of the $n{-}2$ elements $r^{(i)}$ for  $i \in \{1,2,\ldots, n{-}2\}$.

As a free algebra, this product simply consists of concatenating indices
\begin{align}\label{eq: orderingproduct}
    r^{(\alpha)}\cdot r^{(\beta)}:=r^{(\alpha\beta)}\, ,
\end{align}
and in general builds objects with indices consisting of words taken from  the alphabet  $\{1,2,\ldots, n{-}2\}$. It is clear from its definition that this product is non-abelian. For now this is the only structure we will impose on the orderings, however we shall add slightly more in the next section.
The fusion product of the extended generators is defined similarly and is induced through their definition \eqref{eq: extendedGen} in terms of a tensor structure and through the previously defined products \eqref{eq: shuffleproduct} and \eqref{eq: orderingproduct}. It is given by
\begin{align}\label{eq: extendedQS}
\begin{split}
    T^{(\alpha)}_{(\tau_1), \ldots ,(\tau_r)}\star T^{(\beta)}_{(\rho_1), \ldots ,(\rho_s)}&:= (r^{(\alpha)} \otimes  T_{(\tau_1),(\tau_2)\ldots(\tau_r)})\star(r^{(\beta)} \otimes  T_{(\rho_1),(\rho_2)\ldots(\rho_s)})\\
    &= (r^{(\alpha)}\cdot r^{(\beta)}) \otimes  (T_{(\tau_1),(\tau_2)\ldots(\tau_r)}\star T_{(\rho_1),(\rho_2)\ldots(\rho_s)})\\
    &=
   r^{(\alpha\beta)} \otimes \sum_{ \genfrac{}{}{0pt}{}{\sigma\lvert_{\{\tau\}} =\{(\tau_1), \ldots ,(\tau_r)\}}{\sigma\lvert_{\{\rho\}} =\{(\rho_1), \ldots ,(\rho_s)\} } }
    (-1)^{t-r-s} T_{(\sigma_1), \ldots ,(\sigma_t)}\, .
\end{split}
\end{align}
This new product corresponds to the tensor extension of the original product via $\star \mapsto \cdot \otimes \star$, where for convenience we use the same symbol for both.

The actual fusion products we will perform in order to build a BCJ numerator will involve orderings $\alpha$ and $\beta$ of disjoint sets, hence the product $\alpha\beta$ will also be an ordering of their union.

This extension of the quasi-shuffle algebra is non-abelian since the product on the superscripts is non-abelian:
\begin{align}
\begin{split}
     [T^{(\alpha)}_{(\tau_1), \ldots ,(\tau_r)}\, , \, T^{(\beta)}_{(\rho_1), \ldots ,(\rho_s)}]&:= T^{(\alpha)}_{(\tau_1), \ldots ,(\tau_r)}\star T^{(\beta)}_{(\rho_1), \ldots ,(\rho_s)}- T^{(\beta)}_{(\rho_1), \ldots ,(\rho_s)}\star T^{(\alpha)}_{(\tau_1), \ldots ,(\tau_r)}\\
     &=(r^{(\alpha\beta)} - r^{(\beta\alpha)})\otimes (T_{(\tau_1),(\tau_2)\ldots(\tau_r)}\star T_{(\rho_1),(\rho_2)\ldots(\rho_s)})\\ &\neq 0\, .
\end{split}
\end{align}
We can now build algebraic pre-numerators with any ordering $\sigma$ of labels $\{1,2,\ldots,  n{-}2\}$ by taking successive products of extended generators
\begin{equation}
    \widehat\npre(\sigma_{1}\sigma_2 \ldots \sigma_{n{-}2})\coloneqq T_{(\sigma_1)}^{(\sigma_1)}\star T_{(\sigma_2)}^{(\sigma_2)}\star \cdots\star T_{(\sigma_{n{-}2)}}^{(\sigma_{n{-}2})}\,,
\end{equation}
and further we can define algebraic BCJ numerators by simply expanding in terms of algebraic pre-numerators. For example, at  five points one has
\begin{equation}
     \widehat\cN([[1,2], 3])= \widehat\cN(123)-\widehat\cN(213)+\widehat\cN(321)-\widehat\cN(312)\,.
\end{equation}
These algebraic pre-numerators can then be mapped to physical pre-numerators using a map which we again denote by $\langle\bullet\rangle$. If we recall the original definition \eqref{eq: extendedBracket2} of $\langle\bullet\rangle$, we can extend it to arbitrary orderings as follows, 
\begin{align}
\label{eq: extendedBracket2-bis}
         \langle T^{(\alpha)}_{(\tau_1),(\tau_2),\ldots,(\tau_r)} \rangle = \frac{v\Cdot F_{\tau_1}\Cdot V_{\Theta^\alpha(\tau_2)}\Cdot F_{\tau_2}\ldots \Cdot V_{\Theta^\alpha(\tau_r)}\Cdot F_{\tau_r}\Cdot v}{(n-2)v\Cdot p_{\tau_1 [1]} v\Cdot p_{\tau_1}\ldots v\Cdot p_{\tau_1\cdots\tau_{r-1}}}\,,  
    \end{align}
    where now $\Theta^{(\alpha)}(\tau_i)$ is the subset of $\tau_1\cup\cdots\cup\tau_{i-1}$ which are less than the first leg in $\tau_i$ \textit{with respect to the ordering} $\alpha$, for example if $T^{(254163)}_{(13),(25),(46)}$ then $\Theta^{(254163)}(46)=\{2,5\}$. If any of the sets $\Theta^{(\alpha)}(\tau_i)$ happens to be empty then we map that generator to zero. The expression $F_{\tau_j}$ denotes an ordered product of $F_i$ for $i \in \tau_j$ where the $\tau_j$ are also ordered \textit{with respect to the ordering} $\alpha$.

\subsection{The Hopf algebra of orderings}

The extension of the quasi-shuffle algebra described above can also be turned into a Hopf algebra by defining a new coproduct, counit and antipode. This can be achieved by simply tensoring the original Hopf algebra of quasi-shuffles with another Hopf algebra of orderings which acts on the 
$r^{(\alpha)}$, as we now discuss. 

\textbf{The coproduct:} To begin building such a Hopf algebra of orderings, we define a coproduct to the concatenation product \eqref{eq: orderingproduct}. As in the quasi-shuffle case, the coproduct is a map from the space of orderings $\cO$ to the tensor product space $\cO\otimes \cO$.  It turns out that there are many different choices for it. As before, we can define the coproduct by its action on the simplest object, $r^{(i)}$, and then extend to all other elements using linearity and compatibility with the product: $\cop(r^{(\alpha)}\Cdot r^{(\beta)})=\cop(r^{(\alpha)})\Cdot\cop(r^{(\beta)})$.  We will also abuse notation here and use the same symbol for the coproduct,  as in the quasi-shuffle Hopf algebra. 

To explain the possible choices of coproduct we start with the following simple ansatz
\begin{equation}
 \cop_{a,b}(r^{(i)})\coloneqq a(\Id \otimes r^{(i)}+ r^{(i)}\otimes \Id) + b\, r^{(i)}\otimes r^{(i)}\, , 
\end{equation}
where $a$ and $b$ are arbitrary constants at least one of which is non-zero. In order to satisfy co-associativity,
and compatibility with a non-trivial counit and antipode (see \eqref{eq: diagramcounitRel1} and \eqref{eq: diagramantipodedefproperty}), we require the coproduct to be symmetric between the left and right parts of the tensor product. If we consider the case were $b=0$ and $a=1$ the coproduct will not preserve the ordering of legs in the pre-numerator, for example:%
\footnote{Interestingly, this form of the coproduct for the concatenation product \eqref{eq: orderingproduct} appears in \cite{hoffman2000quasi}, where one considers the (quasi-)shuffle algebra as dual to the free algebra. However, we could not find any physical interpretation for it.}
\begin{equation}
    \cop_{a=1,b=0}(r^{(12)})= \Id\otimes r^{(12)}+r^{(1)} \otimes r^{(2)}+r^{(2)} \otimes r^{(1)} + r^{(12)}\otimes \Id\,.
\end{equation}
The whole point of introducing the extended algebra was to make the ordering of particles manifest, so instead we will choose $a=0$ and $b=1$, which gives for a general ordering:
\begin{equation}
    \cop(r^{(\alpha)})\coloneqq \cop_{a=0, b=1}(r^{(\alpha)})= r^{(\alpha)}\otimes r^{(\alpha)}\,.
\end{equation}
It is straightforward to show that this definition also satisfies co-associativity as per diagram \eqref{eq: diagramcoassociativity}.

\textbf{The counit:} The counit can also be simply defined, to yield a bialgebra of the~$r^{(\alpha)}$,
\begin{equation}
    \epsilon(r^{(\alpha)})= \Id\, , \qquad \epsilon(\Id)=\Id\, , 
\end{equation}
where again we are abusing notation and using $\Id$ to label the identity element of the product $\Cdot$ in \eqref{eq: orderingproduct} and the quasi-shuffle product \eqref{eq: stuffleExplicit}. This definition immediately satisfies the defining properties of the counit, the diagrams \eqref{eq: diagramcounitRel1} and \eqref{eq: diagramcounitRel2}.

\textbf{The antipode:} Finally, to construct a Hopf algebra for the orderings $r^{(\alpha)}$ we need to define an antipode $S$, which by definition must satisfy (see diagram \eqref{eq: diagramantipodedefproperty})
\begin{equation}
   \cdot (S\otimes\Id) \cop(r^{(\alpha)})= \cdot (\Id\otimes S) \cop(r^{(\alpha)}) = \epsilon(r^{\alpha})\,.
\end{equation}
Using our choices for the coproduct and the counit, this becomes
\begin{equation}
    S(r^{(\alpha)})\Cdot r^{(\alpha)}= r^{(\alpha)}\Cdot S(r^{(\alpha)})= \Id\,.
\end{equation}
It is clear from the above relation that $S(r^{(\alpha)})$ should correspond to the inverse of $r^{(\alpha)}$, however in the free algebra of the $r^{(\alpha)}$ no such inverses exist. Additionally, the antipode must be an antihomomorphism from the space of $r^{(\alpha)}$ to itself, in particular it must satisfy
\begin{equation}
    S(r^{(ij)})=S(r^{(j)})\Cdot S(r^{(i)})\,.
\end{equation}
To solve both of the above relations we must include inverses and instead consider a \textit{group algebra} of orderings. The antipode will then be identified with the map which sends a group element to its inverse --  this is a very general construction applicable to any group known as a group Hopf algebra.

We began with a free algebra of orderings $r^{(\alpha)}$ and in order to introduce the antipode we require inverse elements. The most general construction we could employ is a group algebra of the free group with $n{-}2$ generators. This group algebra is constructed in a  manner very similar to the free algebra, and consists of linear combinations of products of generators $g$ in the group and inverse elements $g^{-1}$, while assuming \textit{only} the group axioms. However, to define the group algebra which corresponds to orderings we require that the antipode of a particular ordering  $r^{(\alpha)}$
is also an ordering of the same indices contained in $\alpha$. As such we define the inverse of an ordering $\alpha$ to be the reversed ordering $\widebar{\alpha}$:
\begin{equation}
r^{(\alpha)}\Cdot r^{(\beta)}=r^{(\alpha\beta)},  \qquad r^{(\alpha)}\Cdot r^{(\widebar{\alpha})}= \Id\, , \end{equation}
for example $r^{(\widebar{123})}= r^{(321)}$, and as before we can take linear combinations of orderings.  Note that the above conditions imply $r^{(i)}\Cdot r^{(i)}=\Id$. Instead of a free group this is the group freely generated by $n{-}2$ \textit{self-inverse} elements, where ``freely'' here means assuming no other group relations beyond the axioms. This group is known as the {\bf universal Coxeter group} with $n{-}2$ generators.

Now that we have inverses in place,  it is clear that we should define the antipode of an ordering as 
\begin{equation}
    S(r^{(\alpha)})= r^{(\widebar\alpha)}\,.
\end{equation}
This definition also satisfies the other compatibility conditions required by the diagrams~\eqref{eq: diagramAntipodeRelations}.

To summarise, we have defined a coproduct, counit and antipode for the algebra of orderings with product given by \eqref{eq: orderingproduct}. The resulting Hopf algebra has the interesting property of requiring an underlying \textit{group algebra} structure of the universal Coxeter Group such that the antipode acting on an element is identified with the inverse element given by the reversed ordering. This is known as a group Hopf algebra and is a general construction that can be made for any group.

\subsection{Combining the Hopf algebras}
Equipped with a Hopf algebra of orderings, we can now describe the Hopf algebra structure of the non-abelian quasi-shuffle algebra by tensoring the ordering algebra $\cO$ with the quasi-shuffle algebra $\mathfrak{A}$. As advertised at the beginning of this section, and much like the fusion product \eqref{eq: extendedQS} for the extended generators, the coproduct, counit and antipode of the extended algebra are simply given by a product structure.

The coproduct on the extended generators is
\begin{align}\label{eq: extendedcoproduct}
\begin{split}
    \cop(T_{(\tau_1),\ldots, (\tau_r )}^{(\alpha)}):=& (\Id\otimes\mathcal{T}\otimes\Id)\cop(r^{(\alpha)})\otimes \cop(T_{(\tau_1),\ldots, (\tau_r )})\\
    =&\sum_{i=0}^{r}\left(T_{(\tau_1),\ldots, (\tau_i )}^{(\alpha)}\otimes T_{(\tau_{i+1}),\ldots, (\tau_r )}^{(\alpha)}\right)\, ,
\end{split}
\end{align}
where $\mathcal{T} {:}\,\cO\otimes\mathfrak{A}\rightarrow \mathfrak{A}\otimes \mathcal{O}$ is a transposition operator which swaps the entries in the tensor product. It is then an immediate consequence that the coproduct \eqref{eq: extendedcoproduct} is compatible with the product defined in \eqref{eq: extendedQS} and therefore  satisfies the diagram \eqref{eq: diagramcoproductcompatible}. We also again define the reduced coproduct $\Delta'$ which removes all trivial generators without subscripts, i.e.~$T^{(\alpha)}$.

Like the coproduct, the counit is extended such that it acts on both the generators $T_{\omega}$ and the ordering $r^{(\alpha)}$. That is,
\begin{align}
    \epsilon(T^{(\alpha)}_{\omega}):=\epsilon(r^{(\alpha)})\otimes \epsilon(T_{\omega}) =\Id \otimes \epsilon(T_{\omega})= 0\, ,
\end{align}
except for when the element $T_{\omega}=\Id$,  in which case  we have  
\begin{align}
    \epsilon(T^{(\alpha)}):=\epsilon(r^{(\alpha)})\otimes \epsilon(\Id) =\Id \otimes \Id= \Id\, .
\end{align}
Again it is trivial to check that this counit is compatible with the product \eqref{eq: extendedQS} as required by the diagram \eqref{eq: diagramcounitRel2}.

Having defined the coproduct and counit we have fully specified our bialgebra, however to be a Hopf algebra we also need an antipode.
In our case we can  define the antipode acting on extended generators as 
\begin{align}
\begin{split}
S(T^{(\alpha)}_{(\tau_1),\ldots, (\tau_r)})&:=S(r^{(\alpha)})\otimes S(T_{(\tau_1),\ldots, (\tau_r)})=r^{(\overline\alpha)}\otimes\bigg(-\sum_{i=0}^{r-1}S(T_{(\tau_1),\ldots, (\tau_i)})\star T_{(\tau_{i+1}),\ldots,(\tau_r)}\bigg)\, \\
&=\sum_{h=0}^{r-1}(-1)^{h+1}\sum_{r>i_1>i_2\cdots>i_h>1}T^{(\overline\alpha)}_{(\tau_{r}\ldots\tau_{i_1+1}),(\tau_{i_1}\ldots\tau_{i_2+1}),\ldots,(\tau_{i_h}\ldots\tau_{1})}\, ,
\end{split}
\end{align}
where $(\tau_{i}\ldots\tau_{j})$ denotes the union of the $\tau_k$ from indices $i$ to $j$. 
It is again a simple exercise to check the required properties given by diagrams \eqref{eq: diagramantipodedefproperty} and  \eqref{eq: diagramAntipodeRelations}. Acting on an algebraic pre-numerator the antipode reverses the gluon labels up to a factor of $(-1)^{n{-}2}$
\begin{align}\label{eq: antipodeonprenum}
\begin{split}
    S\left(T^{(1)}_{(1)}\star\cdots\star T^{(n{-}2)}_{(n{-}2)}\right)&=S(r^{(12\ldots n{-}2)})\otimes S(T_{(1)}\star\cdots\star T_{(n{-}2)})\\
    &=r^{(n{-}2\ldots21)}\otimes (-1)^{n{-}2}\, T_{(1)}\star\cdots\star T_{(n{-}2)}\\
    &=(-1)^{n{-}2} \widehat\npre(n{-}2\ldots 21 )\, .
\end{split}
\end{align}
It is interesting to note that the antipode acting on an algebraic BCJ numerator simply mirrors the associated cubic graph up to an overall minus sign. For example, the antipode acting on a
left-nested BCJ numerator gives
\begin{align}\label{eq: antipodeonBCJ}
\begin{split}
    S\, \widehat\cN([1,2,3,\ldots,n{-}2])&=(-1)^{n{-}2} \widehat\cN([n{-}2,n{-}3,n{-}4,\ldots,1])\\
    &=-\widehat\cN([1,\ldots,[n{-}4,[n{-}3,n{-}2]]\ldots])\, ,
\end{split}
\end{align}
where the final commutator in \eqref{eq: antipodeonBCJ} is then written in the original order but has gone from left-nested to right-nested, which gives an additional $(-1)^{n-3}$ factor and the mirror of the original cubic graph.

As an example,  consider the case of four gluons, where we have five cubic graphs: a pair of graphs and their mirrors with commutators $\{[[1,2],3],4],\,[1,[2,[3,4]]]\}$ and $\{[[1,[2,3]],4],\,[1,[[2,3],4]]\}$, and the mirror-symmetric graph corresponding to the commutator $[[1,2],[3,4]]$. Under the antipode map we have
\begin{align}
\begin{split}
    S\, \widehat\cN([[1,2],[3,4]]) = -\widehat\cN([[1,2],[3,4]])\,,\\
    S\, \widehat\cN([[1,2],3],4]) = -\widehat\cN([1,[2,[3,4]]])\,,\\
    S\, \widehat\cN([[1,[2,3]],4]) = -\widehat\cN([1,[[2,3],4]])\,.
\end{split}
\end{align}
We can also construct an algebraic amplitude by replacing the kinematic BCJ numerator with the algebraic one, 
\begin{align}\label{eq: algebraicamplitude}
    \widehat A(12\ldots n{-}2)&\, \coloneqq\, \sum_{\commut \in \rho} { \widehat\npre(\commut)\over d_\commut}\,.
\end{align}
The HEFT amplitude \eqref{eq:newDC}  satisfies the reflection identity \eqref{eq:reflrel}, which is inherited from the well-known reflection identity of gluon-scalar amplitudes before taking the heavy-mass limit. We expect the algebraic amplitude to satisfy its own reflection identity, generated by the action of the antipode, which we now describe.

 First, we define an operator $\bar S$ which consists of the antipode $S$ and a reversal of the labels in the massless propagators, hence giving the propagators $d_{\bar\Gamma}$ of the mirrored graph $\bar\Gamma$. Then, under the action of $\bar{S}$, the algebraic amplitude picks up an overall sign
\begin{align}\label{eq: antipodeamplitude}
\begin{split}
    \bar S\, \widehat A(12 \ldots n{-}2) &\coloneqq \sum_{\commut \in \rho} {S \widehat\npre(\commut)\over d_{\widebar{\commut}}}\\
    &= \sum_{\commut \in \rho} {- \widehat\npre(\bar\commut)\over d_{\widebar{\commut}}}\\
    &= -\widehat A(12 \ldots n{-}2)\, ,
\end{split}
\end{align}
 where we have first used  the second line of \eqref{eq: antipodeonBCJ}, and then  the fact that the HEFT amplitude is a sum over all cubic graphs, and hence every BCJ numerator appears together with its mirror. The action of $\bar{S}$ simply multiplies by a factor of $-1$.
 
 The action of $\bar{S}$ on algebraic HEFT  amplitudes can also be written as
\begin{align}\label{eq: antipodeamplitude2}
\begin{split}
     \bar S\, \widehat A(12 \ldots n{-}2) =(-1)^{n{-}2} \widehat A(n{-}2\ldots 21)\, ,
\end{split}
\end{align}
using the first line of \eqref{eq: antipodeonBCJ}. Finally, combining  \eqref{eq: antipodeamplitude} and \eqref{eq: antipodeamplitude2} we have
\begin{equation}
     \widehat A(12 \ldots n{-}2) =(-1)^{n{-}1} \widehat A(n{-}2\ldots 21)\, , 
\end{equation}
which reproduces the reflection identity \eqref{eq:reflrel} for the algebraic HEFT amplitude and relates it to the action of the antipode.\black

In summary, in this section we have  constructed a non-abelian quasi-shuffle Hopf algebra which allows us to treat all possible colour orderings on the same footing, unlike its abelian counterpart that only applied to the
canonical ordering $(1, 2, \ldots, n{-}2)$. As a  bonus we 
were able to give physical interpretations to the various operations of the Hopf algebra at the level of the numerators, rather than the pre-numerator. We will discuss in the next section in detail how the co-product of the new algebra is related to the factorisation of BCJ numerators.

\section{Factorisation behaviour of  BCJ numerators and amplitudes}\label{sec:factorisation2}

In this section we  discuss the factorisation properties of the BCJ numerators and HEFT amplitudes with gluons and gravitons. Just like the pre-numerators, the BCJ numerators and the HEFT amplitudes  are formal polynomials in the  $T_{\omega}$, and we can use the coproduct to characterise the factorisation behaviour. To do this we must define an extension of the replacement rule $\cC$ for the coproduct  defined earlier in \eqref{eq: repRuleC}. This is given by
 \begin{align}
 \label{bigC}
 \begin{split}
       \cC\Big(  T_{\omega_1}^{(\alpha)}\otimes T_{\omega_2}^{(\alpha)}\otimes \cdots\otimes T_{\omega_i}^{(\alpha)}\Big)& =\frac{\|\omega_1 \| \|\omega_2\|\ldots \|\omega_i\|}{\|\omega_1 \|+\|\omega_2\|+\ldots\|\omega_i\|}\frac{1}{v\Cdot p_{\omega_1}v\Cdot p_{\omega_1\omega_2}\cdots v\Cdot p_{\omega_1\cdots \omega_{i-1}}} \\
       &\times\Big( T^{(\alpha|_{\omega_1})}_{\omega_1} p_{\Theta^{(\alpha)}(\omega_2)}\Cdot p_{\omega_2(1)} T^{(\alpha|_{\omega_2})}_{\omega_2} \cdots p_{\Theta^{(\alpha)}(\omega_i)}\Cdot p_{\omega_i(1)} T^{(\alpha|_{\omega_i})}_{\omega_i}\Big)\, ,
       \end{split}
   \end{align}
where we chose to restrict the ordering in the superscript to include only those indices appearing in the subscript.
For example, we have
\begin{align}
\begin{split}
    &\lan T_{(13)}^{(524163)}\otimes T_{(52)(6)}^{(524163)}\otimes T_{(4)}^{(524163)}\ran =0\, , \\
    &  \lan T_{(52)}^{(524163)}\otimes T_{(46)}^{(524163)}\otimes T_{(13)}^{(524163)}\ran =\frac{1}{v\Cdot p_{52} v\Cdot p_{5246}}  T^{(52)}_{(52)} p_{52}\Cdot p_4 T^{(46)}_{(46)} p_{524}\Cdot p_1 T^{(13)}_{(13)}\, .
    \end{split}
\end{align}
With this definition, for each cubic graph or nested commutator structure $\Gamma$, we have
\begin{align}\label{eq: kcopAlgBCJ}
 \Big\langle\Res_{(v\Cdot p_{\omega_1},\ldots v\Cdot p_{\omega_k} )}\,\mathcal{C}\, \cop^{\prime (k)}(\widehat\npre(\Gamma))\Big\rangle=  \Res_{(v\Cdot p_{\omega_1},\ldots v\Cdot p_{\omega_k} )}\, \npre(\Gamma,v)
   \, . 
\end{align}
Finally, before moving on to some examples we note an additional useful relation between left-nested BCJ numerators and the pre-numerator, proven in \cite{Brandhuber:2021bsf}
\begin{align}
\label{pretoBCJ}
    \npre([12\ldots n{-}2])= (n-2)\npre(12\ldots n{-}2)\,.
\end{align}
Note that this relation is only valid at the level of the kinematic pre-numerator and not at the level of the algebraic 
pre-numerator.%
\footnote{We have attempted to implement this relation at the level of the algebra,   for example by requiring that  $\widehat{\npre}([1,2])=2\widehat{\npre}(12)$. However this spoils the definition of the coproduct such that it is no longer well defined.}
\subsection{Five-point example}
To illustrate the factorisation of BCJ numerators we consider some examples, beginning at five points. In this case there are three cubic graphs in the HEFT, and the corresponding BCJ numerators are $\widehat\npre([[1,2],3])$, $\widehat\npre([1,[2,3]])$,  and $\widehat\npre([[1,3],2])$. 
Single-cut factorisations are in one-to-one correspondence with terms in the single coproduct $\cop ( \widehat\npre(\Gamma))$. Using \eqref{bigC}   one finds
\begin{align}
\begin{split}
   \cC \cop^{\prime} ( \widehat\npre([[1,2],3]))&={1\over v\Cdot p_1}T^{(1)}_{(1)}p_1\Cdot p_2\Big({4\over 3}\widehat\npre(23) -{2\over 3}\widehat\npre(32)\Big)\\
   &-{1\over v\Cdot p_2}T^{(2)}_{(2)}p_2\Cdot p_1\Big({4\over 3}\widehat\npre(13) -{2\over 3}\widehat\npre(31)\Big)\\
   &-{2\over 3v\Cdot p_{12}}T^{(3)}_{(3)} \Big((p_1\Cdot p_{3}+p_{12}\Cdot p_{3})\widehat\npre(12)-(p_{12}\Cdot p_{3}+p_2\Cdot p_{3})\widehat\npre(21)\Big)\, .
\end{split}
\end{align}
The single coproduct of the other two cubic graphs are related to the above by swapping the indices and the Jacobi relation
\begin{align}
\begin{split}
      \cC \cop^{\prime} ( \widehat\npre([[1,3],2]))&=\cC \cop^{\prime} ( \widehat\npre([[1,2],3]))\Big|_{2\longleftrightarrow 3}\, ,\\
      \cC \cop^{\prime} ( \widehat\npre([1,[2,3]]))&=\cC \cop^{\prime} ( \widehat\npre([[1,2],3]))-\cC \cop^{\prime} ( \widehat\npre([[1,3],2]))\, .
      \end{split}
\end{align}

Beyond the single-cut factorisation, the (multiple)  coproduct gives terms in correspondence with  multi-factorisation behaviour.
The double-cut factorisation behaviour can be found from the double coproduct as follows:
\begin{align}\label{eq: 5ptdoublecut}
\begin{split}
\cC\cop^{\prime(2)} ( \widehat\npre([[1,2],3]))
&= \Big({p_{1}\Cdot p_{2} p_{12}\Cdot p_3\over v\Cdot p_1 v\Cdot p_{12}}+{p_{1}\Cdot p_{3} p_{1}\Cdot p_2\over v\Cdot p_1 v\Cdot p_{13}}\Big) T^{(1)}_{(1)}T^{(2)}_{(2)}T^{(3)}_{(3)}\, , \\
\cC\cop^{\prime(2)} ( \widehat\npre([[1,3],2]))
&= \Big({p_{1}\Cdot p_{3} p_{13}\Cdot p_2\over v\Cdot p_1 v\Cdot p_{13}}+{p_{1}\Cdot p_{2} p_{1}\Cdot p_3\over v\Cdot p_1 v\Cdot p_{12}}\Big) T^{(1)}_{(1)}T^{(2)}_{(2)}T^{(3)}_{(3)}\, , \\
\cC\cop^{\prime(2)} ( \widehat\npre([1,[2,3]]))
&= \Big({-p_{1}\Cdot p_{3} p_{2}\Cdot p_3\over v\Cdot p_1 v\Cdot p_{13}}+{p_{1}\Cdot p_{2} p_{2}\Cdot p_3\over v\Cdot p_1 v\Cdot p_{12}}\Big) T^{(1)}_{(1)}T^{(2)}_{(2)}T^{(3)}_{(3)}\, .
\end{split}
\end{align}
It is also possible but more complicated to check the factorisation behaviour from the BCJ numerator directly, and doing so we confirm the relation \eqref{eq: kcopAlgBCJ} for the five-point BCJ numerators. 
For example, the double-cut factorisation behaviour is  
\begin{align}
\begin{split}
   \Big\langle\Res_{v\Cdot p_2,v\Cdot p_1}\cC \cop^{\prime(2)}(\widehat\npre(\Gamma)) \Big\rangle &= \Res_{v\Cdot p_2,v\Cdot p_1} \npre(\Gamma,v)\, , \\
   \Big\langle\Res_{v\Cdot p_{1},v\Cdot p_{12}}\cC \cop^{\prime(2)}(\widehat\npre(\Gamma)) \Big\rangle &= \Res_{v\Cdot p_{1},v\Cdot p_{12}} \npre(\Gamma,v)\, , \\
   \Big\langle\Res_{v\Cdot p_{1},v\Cdot p_{13}}\cC \cop^{\prime(2)}(\widehat\npre(\Gamma)) \Big\rangle &= \Res_{v\Cdot p_{1},v\Cdot p_{13}} \npre(\Gamma,v)\, , 
\end{split}
\end{align}
where the $\Gamma\in \{[[1,2],3],[[1,3],2], [1,[2,3]]\}$.  Similarly to the pre-numerator in Section~\ref{sec: Itercoproduct}, only two of the double cuts are independent. 
\black

 Following the factorisation behaviour for  BCJ numerators, the  factorisation behaviour  of amplitudes can also be characterised by the coproduct. For example,  consider the amplitude $A(123, v)$, whose algebraic form is
\begin{align}
    \widehat A(123)={\widehat\npre([[1,2],3])\over p_{12}^2p_{123}^2}+{\widehat\npre([1,[2,3]])\over p_{23}^2p_{123}^2}\, .
\end{align}
Then the single-cut factorisation  on  $v\Cdot p_1=0$ is  
\begin{align}
   \langle \Res_{v\Cdot p_1}\cC\cop^\prime \widehat A(123)\rangle =  {v \Cdot \eps_1 p_1\Cdot p_2 \npre([2,3],v) \over p_{12}^2p_{123}^2}+ {v \Cdot \eps_1 p_1\Cdot p_{23} \npre([2,3],v) \over p_{23}^2p_{123}^2}={1\over 2}v\Cdot \eps_1 A(23,v)\, ,
\end{align}
which is expected from the factorisation behaviour 
on the heavy-mass propagator. On the other hand, the colour-ordered amplitude has no pole in $v\Cdot p_{13}$. Hence, the single cut on $v\Cdot p_2=0$ vanishes,
\begin{align}
   \langle \Res_{v\Cdot p_2}\cC\cop^\prime \widehat A(123)\rangle =  {v \Cdot \eps_2 p_1\Cdot p_2 \npre([1,3],v) \over p_{12}^2p_{123}^2}+ {v \Cdot \eps_2 (p_1\Cdot p_2-p_{13}\Cdot p_{2}) \npre([1,3],v) \over p_{23}^2p_{123}^2}=0\, .
\end{align}
The double cut can be calculated similarly;   we have only one independent double cut  due to the missing pole at $v\Cdot p_2 \!=\!0$. This agrees with the fact that the factorisation of an amplitude does not depend on the ordering in which we take the cuts, in contrast to the pre-numerator discussed in Section~\ref{sec: Itercoproduct}.

\subsection{Six-point example}
We now consider the  six-point case, and mainly focus on the left-nested commutator $[1,2,3,4]$. Other commutators in the DDM basis  are directly obtained by relabelling the indices of the gluons while the algebraic numerators beyond the DDM basis are then derived from the Jacobi relations. 

The BCJ numerator for the left-nested commutator $[1,2,3,4]$ has the following coproduct structure, 
\begin{align}
\begin{split}
   \cC \cop' \widehat\npre([1,2,3,4])&=\frac{2 \widehat\npre(13) \widehat\npre(24) p_1\Cdot p_2}{v\Cdot p_{13}}-\frac{\widehat\npre(31) \widehat\npre(24) p_1\Cdot p_2}{v\Cdot p_{13}}-\frac{\widehat\npre(12) \widehat\npre(43) p_1\Cdot p_3}{v\Cdot p_{12}}\\
   &+\frac{\widehat\npre(12) \widehat\npre(34) \left(2 p_1\Cdot p_3+p_2\Cdot p_3\right)}{v\Cdot p_{12}}-\frac{\widehat\npre(13) \widehat\npre(42) p_1\Cdot p_2}{v\Cdot p_{13}}\\
   &+{3\over 4}\Big(\frac{\widehat\npre(4) \widehat\npre(123) \left(p_{23}\Cdot p_4+2 p_1\Cdot p_4\right)}{v\Cdot p_{123}}+\frac{\widehat\npre(3) \widehat\npre(412) p_{12}\Cdot p_3}{v\Cdot p_3}\\
   &+\frac{ \widehat\npre(1) (2 \widehat\npre(234)+\widehat\npre(432)-\widehat\npre(324)-\widehat\npre(423)) p_1\Cdot p_2}{v\Cdot p_1}\\
   &-\frac{\widehat\npre(3) \widehat\npre(124) \left(2 p_1\Cdot p_3+p_2\Cdot p_3\right)}{v\Cdot p_3}-\frac{\widehat\npre(4) \widehat\npre(312) \left(p_{12}\Cdot p_4+2 p_3\Cdot p_4\right)}{v\Cdot p_{123}}\Big)\\
   &-(1\leftrightarrow 2)\, .
\end{split}
\end{align}
After mapping to kinematics we can find all the factorisation channels from \\
$\langle\Res_{v\Cdot P}\cC \cop \widehat\npre([1,2,3,4])\rangle$, as tabulated below:
\begin{align}
\nn
\begin{array}{|c||c||c|}
\hline
\rowcolor{blue!20}
v\Cdot p_1 & v\Cdot p_{134} &v\Cdot p_{124}\\
\hline\hline
\npre(1,v)\npre([234],v) p_1\Cdot p_2 & \npre(2,v) \npre([134],v) p_1\Cdot p_2 &
\npre(3,v)  \npre([124],v) p_{12}\Cdot p_3\\
  \hline
\rowcolor{blue!20}
v\Cdot p_{14} & v\Cdot p_{13} &v\Cdot p_{12}\\
\hline\hline
 \npre([14],v)\npre([23],v) p_1\Cdot p_2 & \npre([13],v) \npre([24],v) p_1\Cdot p_2 &
\npre([12],v) \npre([34],v) p_{12}\Cdot p_3\\
  \hline
\rowcolor{blue!20}
 &  v\Cdot p_{123}&\\
\hline\hline
& \npre(4,v)\npre([123],v) p_{123}\Cdot p_4  &
\\ 
  \hline
\end{array}
\end{align}
For the colour-ordered amplitude, we have 
\begin{align}
\begin{split}
    \widehat A(1234,v)
    &={\widehat\npre([1,2,3,4],v)\over p_{12}^2p_{123}^2p_{1234}^2}+{\widehat\npre([[1,2],[3,4]],v)\over p_{12}^2p_{34}^2p_{1234}^2}-{\widehat\npre([2,3,4,1],v)\over p_{23}^2p_{234}^2p_{1234}^2}\\
    &+{\widehat\npre([3,4,2,1],v)\over p_{34}^2p_{234}^2p_{1234}^2}-{\widehat\npre( [2,3,1,4],v)\over p_{23}^2p_{123}^2p_{1234}^2}\, .
\end{split}
\end{align}
The factorisation behaviour of the amplitude can also be obtained from the coproduct by using the factorisation behaviour of the BCJ numerators. Considering the pole at $v\Cdot p_1=0$ which is present in the colour-ordered amplitude, we have 
\begin{align}
\begin{split}
   \langle \Res_{v\Cdot p_1}\cC\cop \widehat A(1234,v)\rangle &=\Big( {p_1\Cdot p_2 \over p_{12}^2p_{123}^2p_{1234}^2}+{p_1\Cdot p_{234} \over p_{23}^2p_{234}^2p_{1234}^2}+{p_1\Cdot p_{23} \over p_{23}^2p_{123}^2p_{1234}^2}\Big)\npre(1,v)\npre([234],v)\\
   &+ \Big( {p_1\Cdot p_2 \over p_{12}^2p_{34}^2p_{1234}^2}+{p_1\Cdot p_{234} \over p_{34}^2p_{234}^2p_{1234}^2}\Big)\npre(1,v)\npre([2,[3,4]]
   \\
   &={1\over 2} A(1,v) A(234,v)\, .
    \end{split}
\end{align}
On the other hand the colour-ordered amplitude has not pole at $v\Cdot p_2\!=\!0$, which is
consistent with the fact that the corresponding residue vanishes, 
\begin{align}
\begin{split}
  & \langle \Res_{v\Cdot p_2}\cC\cop \widehat A(1234,v)\rangle \ =\Big( -{p_1\Cdot p_2 \over p_{12}^2p_{123}^2p_{1234}^2}+{p_2\Cdot p_{3} \over p_{23}^2p_{123}^2p_{1234}^2}\Big)\npre(2,v)\npre([134],v)\\
   &+\Big( -{p_1\Cdot p_2 \over p_{12}^2p_{34}^2p_{1234}^2}+{p_2\Cdot p_{3} \over p_{23}^2p_{234}^2p_{1234}^2}+{p_2\Cdot p_{34} \over p_{34}^2p_{234}^2p_{1234}^2}\Big)\npre(2,v)\npre([1,[3,4]],v)\\
   &=0\, .
    \end{split}
\end{align}
The factorisation behaviour in the remaining channels is similar and the non-vanishing poles are at $v\Cdot p_1\!=\!0$, $v\Cdot p_{12}\!=\!0$,  and $v\Cdot p_{123}\!=\!0$. It is straightforward to check that there are no ``composite poles'' (in the sense of \cite{Arkani-Hamed:2009ljj})
and hence the multi-cut behaviour of the amplitude is also unique. This property also holds for general multiplicity amplitudes as expected.

\section{Conclusions and outlook}\label{sec: conclusion}

In this paper we have extended the work presented in its prequel \cite{Brandhuber:2021eyq}, further clarifying the role of the Hopf algebra in the construction of BCJ numerators in the HEFT. In particular we have established several novel connections between the rich algebraic structure of the quasi-shuffle Hopf algebra and fundamental properties of HEFT BCJ numerators. After proving the double copy in the heavy-mass limit, we found a novel relation between the action of the coproduct on generators of the Hopf algebra and factorisation of pre-numerators in the HEFT, where both serve to break up pre-numerators into their building blocks.

As well as linking factorisation to the action of the coproduct, we have also extended the Hopf algebra to a non-abelian version. This was motivated by the non-commutative nature of gluon labels in the physical pre-numerator, which allows the construction of BCJ numerators by considering commutators of pre-numerators with their labels swapped. We exactly emulated this construction of BCJ numerators in a purely algebraic manner by defining a non-abelian quasi-shuffle algebra. Then we constructed the non-abelian algebra in a simple way by tensoring the original quasi-shuffle algebra with an algebra of orderings of the gluon labels. The latter is also a Hopf algebra and is given by the group Hopf algebra of the universal Coxeter group. This non-abelian algebra has the appealing property that the action of the antipode is related to the reversal of the gluon ordering in a numerator, and in turn plays a role in the reflection identity \eqref{eq: HEFTampreversal} of the HEFT amplitude. Finally, after defining a direct algebraic construction of HEFT BCJ numerators, we have related their factorisation properties to the coproduct of the non-abelian quasi-shuffle Hopf algebra.

The results of this paper leave us with several interesting  avenues of research, some of which we now outline.  First, an important question is to assess   the universality of the Hopf algebra. The first indication of this appears in  \cite{Chen:2022nei}, where it was found that very similar Hopf algebra structures govern the colour-kinematics duality in Yang-Mills-scalar theory with a bi-adjoint $\phi^3$ interaction. The construction was also applied  both to scattering amplitudes and form factors with an arbitrary number  of gluons and massive scalars. The results of that paper  point to  the    universality of the Hopf algebra,  
beyond the leading HEFT expansion. The connections between physical properties of amplitudes and mathematical structures of the Hopf algebra  revealed in this paper should have an interesting application in those more general contexts.

One can also consider amplitudes with gravitons and massive particles such as fermions,  vectors and even higher-spin particles, which enter the computation of the  deflection  angle, waveform and metric in the gravitational scattering of Kerr black holes. We expect the colour-kinematics duality and the  double copy  to play a significant role here, and it would be fascinating to  identify  a  kinematic Hopf algebra in this context.

Another important direction concerns  loop  amplitudes.  The colour-kinematics duality and double copy in  Yang-Mills and gravity  appear to hold also at loop level (see for instance \cite{Bern:2019prr}), and it would be very interesting to identify algebraic structures also in loop  amplitudes, especially the loop integrands.

Finally, we are still missing an explicit realisation of the algebra in terms of differential operators, similarly to the work of \cite{Monteiro:2014cda} for the self-dual sector of Yang-Mills and gravity. This would avoid the appearance of abstract generators and the angle-bracket map. 
We leave these and other fascinating questions for future investigation.

\section*{Acknowledgements}

We would like to thank Massimo Bianchi, Sanjaye Ramgoolam  and Oliver Schlotterer for  useful discussions. 
AB, GB, JG and  GT would like to thank the Kavli Institute for Theoretical Physics at the University of California, Santa Barbara, where their research was supported in part by the National Science Foundation under Grant 
No.~PHY-1748958.
This work was supported by the Science and Technology Facilities Council (STFC) Consolidated Grants ST/P000754/1 \textit{``String theory, gauge theory \& duality''} and  ST/T000686/1 \textit{``Amplitudes, strings  \& duality''}, by the European Union's Horizon 2020 research and innovation programme under the Marie Sk\l{}odowska-Curie grant agreement No.~764850 {\it ``\href{https://sagex.org}{SAGEX}''}, and 
 by the National Science Foundation under Grant No. NSF PHY-1748958.
CW is supported by a Royal Society University Research Fellowship No.~UF160350.
The work of GRB and JG  is supported by an STFC quota studentship. No new data were generated or analysed during this study.

\newpage
\appendix

\section{Recursive definition of the quasi-shuffle product} \label{app: RecursiveDefs}
 In this appendix we detail,  for completeness,  a recursive definition  for the usual quasi-shuffle product \cite{hoffman2000quasi,hoffman2017quasi}. It turns out that explicitly writing the generators $T_{\omega}$ leads to clumsy looking formulas, so instead we will write relations involving only the words $\omega = (\tau_1),\ldots , (\tau_r)$. The extension to the full generators is then immediate.
 
 If we have two generators $T_{(\tau_1),\ldots,(\tau_r)}$ and $T_{(\rho_1),\ldots,(\rho_s)}$ then we define the quasi-shuffle product through the following recursive relation 
\begin{align}
\label{eq: stuffleRecursive}
\begin{split}
   &  (\tau_1)\cdots(\tau_r) \star (\rho_1)\cdots(\rho_s) \coloneqq\\
     &(\tau_1)\Big[(\tau_2)\cdots(\tau_r)\star (\rho_1)\cdots(\rho_s)\Big] + 
     (\rho_1)\Big[(\tau_1)\cdots(\tau_r)\star (\rho_2)\cdots(\rho_s)\Big]\\
     &- (\tau_1 \rho_1) \Big[(\tau_2)\cdots(\tau_r)\star(\rho_2)\cdots(\rho_s)\Big]\,. 
\end{split}
\end{align}
 In order for this recursion to eventually end, we need to define an ``empty'' identity element $\Id$ for the quasi-shuffle product~$\star$
\begin{align}
\begin{split}
    \Id\,\, (\tau_1)\cdots(\tau_r) &=(\tau_1)\cdots(\tau_r)\,\, \Id  =(\tau_1)\cdots(\tau_r) \qquad\, \text{``empty'' property}\,,\\[1ex] 
    \Id \star (\tau_1)\cdots(\tau_r) &= (\tau_1)\cdots(\tau_r)\star \Id = (\tau_1)\cdots(\tau_r) \qquad \text{Identity for } \star\, .
\end{split}
\end{align}

\section{Quasi-shuffles from shuffles}
\label{app:B}
One important question is whether or not the quasi-shuffle algebra used to find pre-numerators is unique. For example, can one redefine the generators $T_{(\tau_1)\ldots }$, and change the quasi-shuffle algebra into another algebra with a different product? The answer is that one can indeed define new generators such that we obtain a shuffle algebra instead, although at the expense of making the mapping $\langle \bullet\rangle$ \eqref{eq: extendedBracket2} much more complicated. To do this it will be necessary to introduce some extra machinery. Namely, we will need to define the action of analytic functions on abstract letters and words which will form maps between algebras. In this appendix we will briefly introduce shuffle products before explicitly relating them to quasi-shuffles. Here we apply the methods of \cite{hoffman2017quasi,hoffman2000quasi} to our context, as well as adopting their notation. For example, once again we drop the $T_{\omega}$ notation and initially work with just the words $\omega = (\tau_1),\ldots , (\tau_r)$.
\subsection{The shuffle product}
The shuffle product $\shuffle$ of two words ${(\tau_1)\ldots(\tau_r)}$ and ${(\rho_1)\ldots(\rho_s)}$ can also be defined using a recursive definition
\begin{align}
\label{eq: shuffleRecursive}
\begin{split}
   &  (\tau_1)\cdots(\tau_r) \shuffle (\rho_1)\cdots(\rho_s) \coloneqq\\
     &(\tau_1)\Big[(\tau_2)\cdots(\tau_r)\shuffle (\rho_1)\cdots(\rho_s)\Big] + 
     (\rho_1)\Big[(\tau_1)\cdots(\tau_r)\shuffle (\rho_2)\cdots(\rho_s)\Big]\,.
\end{split}
\end{align}
This is very similar to the recursive definition of the quasi-shuffle given in appendix \ref{app: RecursiveDefs} except that the final ``stuffing'' term $ (\tau_1 \rho_1) [(\tau_2)\cdots(\tau_r)\star(\rho_2)\cdots(\rho_s)]$ in \eqref{eq: stuffleRecursive} is dropped. Like the quasi-shuffle product, the shuffle product is abelian and  there is an ``empty'' identity element $\Id$. 

As an example consider the shuffle products 
\begin{align}
    &(1)\shuffle (2)= (1)(2)+(2)(1)\,,\nn\\
    &(1)\shuffle (2)(3)(4) = (1)(2)(3)(4)+(2)(1)(3)(4)+(2)(3)(1)(4)+(2)(3)(4)(1)\,,
\end{align}
which should be compared to their quasi-shuffle cousins \eqref{eq: shuffleproduct} and \eqref{eq: shuffleproduct2}. Here we can see explicitly that terms like (12) etc. are not included in the shuffle product.

\subsection{Compositions of words}
To define the action of functions on words we first need to introduce the notion of a composition. Given a word 
of length $n$ we define the set of compositions as follows 
\begin{equation}
    \mathfrak{C}(n) := \{(i_1, i_2, \ldots, i_k)\, :\, i_j\in\mathbb{Z}^+, \, i_1+i_2+\cdots + i_k = n\}\, .
\end{equation}
Therefore, a given composition is a list of positive integers which add up to give the length of the word.
These compositions can then act on words to obtain shorter words. Given $I=(i_1,i_2,\ldots , i_k)$ and $\omega = (\tau_1)(\tau_2)\ldots (\tau_n)$ we define 
\begin{equation}
\begin{split}
    I[\omega]  
    &=(\tau_1\ldots\tau_{i_1})(\tau_{i_1+1}\ldots\tau_{i_1+i_2})\ldots (\tau_{i_1+i_2+\cdots +i_{k-1}+1}\ldots\tau_{n})\,. 
\end{split}
\end{equation}
For example, the composition $I = (2,3,1)$ acting on $\omega =(\tau_1)(\tau_2)(\tau_3)(\tau_4)(\tau_5)(\tau_6)$ is
\begin{equation}
    I[\omega] = (\tau_1\tau_2)(\tau_3\tau_4\tau_5)(\tau_6) \, .
\end{equation}
\subsection{Functions of words}
To map the quasi-shuffle algebra to a different algebra we will need to define the action of analytic functions on words. Much like defining the action of a function on a matrix; the action of a function on words is defined using its Taylor series. Given a function of a single variable $f(x)$ which is analytic and zero at $x=0$ we write its Taylor series as
\begin{equation}
    f(x) = \sum_{i=1}^\infty a_i x^i\, .
\end{equation}
Then we define the corresponding \textit{linear} function $\Psi_f$ on a word $\omega$ using compositions \cite{hoffman2000quasi,hoffman2017quasi} 
\begin{equation}
    \Psi_f (\omega) = \sum_{(i_1,\ldots ,i_k)\in\ \mathfrak{C}(\mathit{l}(\omega))} 
    a_{i_1}\ldots a_{i_k}\, (i_1,\ldots ,i_k)[\omega]\, ,
\end{equation}
where $l(\omega)$ is the length of the word $\omega$ i.e. the number of letters it contains.
This definition satisfies intuitive properties when taking compositions and inverses of functions 
\begin{equation}
    \Psi_f  \Psi_g = \Psi_{f\circ g}, \quad \Psi_{f^{-1}}= (\Psi_f)^{-1}\,.
\end{equation}
Of particular importance for us are the exponential and logarithm maps which take us between the shuffle and quasi-shuffle algebra\footnote{Note that due to linearity we have $\exp(\omega_1+\omega_2)= \exp(\omega_1)+ \exp(\omega_2)$ and $\log(\omega_1+\omega_2)= \log(\omega_1)+ \log(\omega_2)$. }
\begin{equation}
    \begin{split}
         \exp(\omega) :=&\, \Psi_{e^x-1}(\omega)=  \sum_{(i_1,\ldots, i_k)\in \mathfrak{C}(\mathit{l}(\omega))} \frac{1}{i_1,\ldots ,i_k!} (i_1,\ldots ,i_k)[\omega]\, ,\\
         \log(\omega) :=&\, \Psi_{\log(1+x)}(\omega)=  \sum_{(i_1,\ldots ,i_k)\in \mathfrak{C}(\mathit{l}(\omega))} \frac{(-1)^{\mathit{l}(\omega)-k}}{i_1,\ldots ,i_k} (i_1,\ldots ,i_k)[\omega]\,.\\
    \end{split}
\end{equation}
 In addition we will need a function which introduces a sign in our version of the quasi-shuffle algebra \cite{hoffman2000quasi,hoffman2017quasi}%
 \footnote{In \cite{hoffman2000quasi,hoffman2017quasi} this function is denoted by $T$, however here we instead use $\chi$ for obvious reasons. }
\begin{equation}
    \chi(\omega) := \Psi_{-x}(\omega)= (-1)^{\mathit{l}(\omega)}\omega \,.
\end{equation}
With these functions in hand, we can map shuffles to quasi-shuffles and vice versa as follows, for any words $\omega_1$ and $\omega_2$ we have 
\begin{equation}\label{eq: expShuffToStuff}
    \begin{split}
         \chi \exp \chi (\omega_1 \shuffle \omega_2) &= \chi\exp \chi (\omega_1) \star \chi\exp \chi (\omega_2)\,,\\
        \chi \log \chi (\omega_1 \star \omega_2) &= \chi\log \chi (\omega_1) \shuffle \chi\log \chi (\omega_2)\,.
    \end{split}
\end{equation}
The functions $ \chi \exp \chi$ and $\chi \log \chi$  are clearly mutual inverses, and thus both are algebra isomorphisms. The equations above can be naturally extended to arbitrary chains of shuffles and quasi-shuffles. In particular, for the pre-numerator we can write (re-introducing the generator notation $T_{\omega}$ here)
\begin{equation}
     \chi\exp \chi(T_{(1)}\shuffle\cdots \shuffle  T_{(n)}) =   \chi\exp \chi(T_{(1)})\star\cdots \star    \chi\exp \chi(T_{(n)}) = T_{(1)} \star \cdots \star  T_{(n)}\,.
\end{equation}
The above relation suggests that we can rebuild the pre-numerator using only shuffles, but to do so we will need to define some new generators as follows
\begin{equation}\label{eq: shuffleGens}
    \widetilde{T}_{\omega}\coloneqq \chi \exp \chi (T_{\omega})= \Psi_{1-e^{-x}}(T_{\omega})\, .
\end{equation}
We will call these generators $\widetilde{T}$ the shuffle generators. It is worth going through a few examples of these new generators to see explicitly how they relate to the original ones:
\begin{equation}
    \begin{split}
    \widetilde{T}_{(i)} &= T_{(i)}\quad\quad \quad \text{for any $i=1,\ldots ,n$}\,,\\
    \widetilde{T}_{(1),(2)} &= T_{(1),(2)} - \frac{1}{2}T_{(12)}\,,\\
    \widetilde{T}_{(1),(2),(3)} &= T_{(1),(2),(3)} - \frac{1}{2}T_{(12),(3)}
     - \frac{1}{2}T_{(1),(23)} + \frac{1}{6}T_{(123)}\,.\\
    \end{split}
\end{equation}
 Note that the shuffle generators are only labelled by brackets containing a single leg like $\widetilde{T}_{(1),(2),(3)}$ etc.
 
 To show the equivalence of these generators to the original quasi-shuffle generators we need to define a shuffle product $\widetilde{\shuffle}$ for the generators $\widetilde{T}$. This product is defined, as expected, by the shuffling of the brackets associated to each generator $\widetilde{T}$
\begin{equation}
    \widetilde{T}_{(i_1),(i_2),\ldots ,(i_r)}\, \widetilde{\shuffle}\, \widetilde{T}_{(j_1),(j_2),\ldots ,(j_s)} \coloneqq  \sum_{(k_1),\ldots , (k_{r+s})\in(i_1),\ldots , (i_r) \shuffle (j_1),\ldots ,(j_s)} \widetilde{T}_{(k_{1}),\ldots , (k_{r+s})}\, ,
\end{equation}
where we are abusing notation and using $(i_1),\ldots , (i_r) \shuffle (j_1),\ldots ,(j_s)$ to refer to the set of terms appearing in the shuffle product. For example we have
\begin{align}
    &\widetilde{T}_{(1)}\widetilde{\shuffle}\widetilde{T}_{(2)}= \widetilde{T}_{(1),(2)}+\widetilde{T}_{(2),(1)}\,,\nn\\
    &\widetilde{T}_{(1)}\widetilde{\shuffle}\widetilde{T}_{(2),(3),(4)}= \widetilde{T}_{(1),(2),(3),(4)}+\widetilde{T}_{(2),(1),(3),(4)}+\widetilde{T}_{(2),(3),(1),(4)}+\widetilde{T}_{(2),(3),(4),(1)}\,.
\end{align}
Using the above definition of $\widetilde{\shuffle}$ and the definition of the shuffle generators \eqref{eq: shuffleGens} we obtain the following relation
\begin{align}
\begin{split}
    \widetilde{T}_{(i_1),(i_2),\ldots , (i_r)}\,\widetilde{\shuffle}\, \widetilde{T}_{(j_1),(j_2),\ldots , (j_s)} 
    &=  \sum_{(k_1),\ldots , (k_{r+s})\in(i_1),\ldots , (i_r) \shuffle (j_1),\ldots ,(j_s)} \widetilde{T}_{(k_{1}),\ldots , (k_{r+s})}\\
    &=  \sum_{(k_1),\ldots , (k_{r+s})\in(i_1),\ldots , (i_r) \shuffle (j_1),\ldots ,(j_s)} \chi \exp \chi (T_{(k_{1}),\ldots , (k_{n})})\\
    &=  \chi \exp \chi(T_{(i_1),(i_2),\ldots , (i_r)}\,\shuffle\, T_{(j_1),(j_2),\ldots , (j_s)})\\
    &=  \chi \exp \chi(T_{(i_1),(i_2),\ldots , (i_r)})\star \chi \exp \chi( T_{(j_1),(j_2),\ldots , (j_s)})\,,
\end{split}
\end{align}
where in the last equality we have used \eqref{eq: expShuffToStuff}. Using this relation, we can finally show how to construct the algebraic pre-numerator out of shuffles
\begin{align}
    \widetilde{T}_{(1)}\widetilde{\shuffle}  \widetilde{T}_{(2)}\widetilde{\shuffle}\ldots\widetilde{\shuffle}\widetilde{T}_{(n-2)}&= \chi \exp \chi(T_{(1)})\star \chi \exp \chi( T_{(2)})\star\ldots\star \chi \exp \chi( T_{(n-2)})\nn\\
    &= T_{(1)}\star T_{(2)}\star \ldots\star T_{(n-2)}\,.
\end{align}
The final step to reconstruct the pre-numerator is to work out a mapping from abstract generators to kinematic quantities which we again denote by $\langle\bullet \rangle$. Conveniently, we can derive such a map simply using the  definition of the shuffle generators $\widetilde{T}$ in terms of the quasi-shuffle generators $T$
\begin{equation}
    \langle \widetilde{T}_{(i_1),\ldots , (i_k)} \rangle \coloneqq \langle \chi \exp \chi (T_{(i_1),\ldots , (i_k)}) \rangle\, ,
\end{equation}
and the quasi-shuffle map defined in \eqref{eq: anglebracketmap}. Here are some explicit examples
\begin{align}
\begin{split}
    \langle \widetilde{T}_{(i)} \rangle &= \langle T_{(i)} \rangle = v\Cdot \eps_{i}\, ,\\
    \langle \widetilde{T}_{(1),(2)} \rangle &= \langle 
    {T}_{(1),(2)} \rangle - \frac{1}{2} \langle T_{(12)}\rangle = -\frac{1}{2} \frac{v\cdot F_1 \Cdot F_2 \Cdot v}{2 v\Cdot p_1}\, ,\\
    \langle \widetilde{T}_{(1),(2),(3)}\rangle &=  \langle T_{(1),(2),(3)}\rangle -\frac{1}{2}\langle T_{(12),(3)}\rangle
     -\frac{1}{2}\langle T_{(1),(23)}\rangle + \frac{1}{6}\langle T_{(123)}\rangle\\
     & =  -\frac{1}{2}\frac{v\Cdot F_1 \Cdot F_2 \Cdot V_{12} \Cdot F_3 \Cdot v}{3 v\Cdot p_1 v \Cdot p_{12}} +\frac{1}{6} \frac{v\Cdot F_1 \Cdot F_2 \Cdot F_3 \Cdot v}{3 v\Cdot p_1} \, .
\end{split}
\end{align}
The point to note is that this map is more complicated than the original one, in that it involves a sum of many terms with no clear simplification. This is not too surprising: we have simplified the algebra to a shuffle algebra, and it is then reasonable to expect that the mapping is more complex. Despite this, one might hope that the angle bracket map $\langle\bullet \rangle$ above could be simplified in some way, or that a totally different angle map could give rise to the same pre-numerators. Alternatively, there could be another choice of pre-numerator altogether, better suited to the shuffle product. Unfortunately, these options seem to be as difficult as finding the original map $\langle\bullet \rangle$ for the quasi-shuffle generators $T$, and in this work we have been unable to find a simple equivalent shuffle version. Indeed, the algebraic pre-numerator constructed from shuffles (for example $\widetilde{T}_{(1)}\widetilde{\shuffle}\widetilde{T}_{(2)}$) contains explicitly fewer terms than the equivalent quasi-shuffle pre-numerator (for example ${T}_{(1)}\star T_{(2)}$) hence the map $\langle\bullet\rangle$ is likely to be more involved.



\section{Pre-numerators with manifest antipodal symmetry}\label{ap: antipodePreNum}

As mentioned in Section~\ref{sec:3}, there is some freedom in constructing  pre-numerators, which can be used  to make certain properties manifest.
For instance,  using the angle-bracket map defined in the main text  creates a pre-numerator which, by construction, makes the crossing symmetry of the gluons manifest. 
As another example of this freedom, in this appendix  we will define a pre-numerator which only picks up a sign under the action of the antipode map -- we will refer to this property as ``antipodal symmetry''.   Hence, these alternative pre-numerators have the same properties under the action of the antipode as the amplitudes, which implies that they obey their own version of the reflection identity \eqref{eq:reflrel}. We  now describe these alternative pre-numerators and their properties in detail. 

We start from the algebraic pre-numerator
\begin{align}
    \widehat \npre(\alpha)= T^{(\alpha_1)}_{(\alpha_1)}\star T^{(\alpha_2)}_{(\alpha_2)} \star \cdots \star  T^{(\alpha_{n-2})}_{(\alpha_{n-2})}\, ,
\end{align}
and the new pre-numerator is defined in terms of the new map $\langle \bullet \rangle_S$ 
\begin{align}
    \npre'(\alpha,v)\coloneqq \langle \widehat \npre(\alpha) \rangle_S, 
\end{align}
 where the new angle-bracket map is defined as 
\begin{equation}
         \langle T^{(\alpha)}_{(\tau_1),(\tau_2),\ldots,(\tau_r)} \rangle_S \coloneqq
         \begin{cases}
            \frac{v\Cdot F_{\tau_1}\Cdot V_{\Theta^\alpha(\tau_2)}\Cdot F_{\tau_2}\ldots \Cdot V_{\Theta^\alpha(\tau_r)}\Cdot F_{\tau_r}\Cdot v}{2v\Cdot p_{1} v\Cdot p_{\tau_1}\ldots v\Cdot p_{\tau_1\cdots\tau_{r-1}}}\, &\text{if $\alpha_1=1$},\\
           - \frac{v\Cdot F_{\tau_r}\Cdot \widebar V_{\widebar\Theta^\alpha(\tau_r)}\ldots \Cdot F_{\tau_2}\Cdot \widebar V_{\widebar\Theta^\alpha(\tau_2)}\Cdot F_{\tau_1}\Cdot v}{2 v\Cdot p_{\tau_1\cdots\tau_{r-1}}\ldots  v\Cdot p_{\tau_1} v\Cdot p_{1}}\, &\text{if $\alpha_{n-2}=1$} \\
           0\, &\text{otherwise},
         \end{cases}  ,
\end{equation}
 where $V_\tau^{\mu\nu}$ and $\Theta^{(\alpha)}(\tau_i)$ are defined in 
 Sections~\ref{sec: MapToKinematics} and \ref{sec:5.1}, respectively, 
 and $\widebar V_\tau^{\mu\nu}= p_{\tau}^\mu v^\nu $  and $\widebar\Theta^{(\alpha)}(\tau_i)$ is the subset of $\tau_1\cup\cdots\cup\tau_{i-1}$ which are larger than the last index in $\tau_i$ \textit{with respect to the ordering} $\alpha$. Note that the $\tau_j$ are also ordered with respect to $\alpha$. For example, we write $T^{(254631)}_{(31),(25),(46)}$ and then since the last index in the superscript is one we use the $\widebar\Theta$ functions which for this example gives:  $\widebar\Theta^{(254631)}(46)=\{31\}$, $\widebar\Theta^{(254631)}(31)=\emptyset$ etc.
As before, if any of the sets $\Theta^{(\alpha)}(\tau_i), \widebar\Theta^{(\alpha)}(\tau_i)$ happen to be empty then we map that generator to zero. Finally, the three-point case is special and as usual we define it separately as
\begin{align}
    \langle T_{(i)}\rangle_S \coloneqq v\Cdot \eps_i \, . 
\end{align}
As  anticipated, the new pre-numerators now obey the antipodal symmetry, 
\begin{equation}
   \langle S \widehat\npre(\alpha)\rangle_S= -\langle  \widehat\npre(\alpha)\rangle_S,
\end{equation}
which, using \eqref{eq: antipodeonprenum}, implies a reflection identity for the alternative pre-numerators which is the same as that obeyed by the amplitude
\begin{equation}
   \langle \widehat\npre(\overline\alpha)\rangle_S= (-1)^{n-1}\langle  \widehat\npre(\alpha)\rangle_S\,,
\end{equation}
where as usual $\overline\alpha$ is the reverse of the labels $\alpha$. \black
For example, at four and five points we have
\begin{align}
\begin{split}
    \langle S\widehat\npre(12)\rangle_S=&\langle T^{(2)}_{(2)}\star  T^{(1)}_{(1)}\rangle_S=-\langle T^{(21)}_{(21)}\rangle_S={v\Cdot F_{21}\Cdot v \over 2v\Cdot p_1}\\
    =&{v\Cdot F_{12}\Cdot v \over 2v\Cdot p_1}=-\langle \widehat\npre(12)\rangle_S\, , \\
     \langle S\widehat\npre(123)\rangle_S=&-\langle T^{(3)}_{(3)}\star T^{(2)}_{(2)}\star  T^{(1)}_{(1)}\rangle_S=\langle T^{(321)}_{(321)}\rangle_S+\cdots\\
    =&-{v\Cdot F_{321}\Cdot v \over 2v\Cdot p_1}+\cdots ={v\Cdot F_{123}\Cdot v \over 2v\Cdot p_1}+ \cdots= -\langle \widehat\npre(123)\rangle_S\, .
\end{split}
\end{align}
Finally,  even though the pre-numerators are different, the BCJ numerators are the same as those obtained from  the map discussed in the main text, namely
\begin{align}
    \langle \widehat\npre(\Gamma)\rangle=\langle \widehat\npre(\Gamma)\rangle_S =\npre(\Gamma, v)\,.
\end{align}
It is useful to  consider a simple example of constructing  BCJ numerators using the new map: 
\begin{align}
\begin{split}
    \langle\widehat\npre([[1,2],3])\rangle_S
    &=\langle\widehat\npre(123)\rangle_S-\cancelto{0}{\langle\widehat\npre(213)\rangle_S}-\cancelto{0}{\langle\widehat\npre(312)\rangle_S}+\langle\widehat\npre(321)\rangle_S\\
    &=-\frac{v\Cdot F_1 \Cdot F_2 \Cdot V_{12}\Cdot F_{3}\Cdot v}{2 v\Cdot p_1 v\Cdot p_{12}} -  \frac{v\Cdot F_1 \Cdot F_3 \Cdot V_{1}\Cdot F_{2}\Cdot v}{2 v\Cdot p_1 v\Cdot p_{13}} + \frac{v\Cdot F_1 \Cdot F_2 \Cdot F_3\Cdot v}{2 v\Cdot p_1}\\
    &-\Big(-\frac{v\Cdot F_3\Cdot\widebar V_{12}\Cdot F_2 \Cdot F_1\Cdot v}{2 v\Cdot p_1 v\Cdot p_{12}} -\frac{v\Cdot F_2\Cdot\widebar V_{1}\Cdot F_3 \Cdot F_1\Cdot v}{2 v\Cdot p_1 v\Cdot  p_{31} } + \frac{v\Cdot F_3 \Cdot F_2 \Cdot F_1\Cdot v}{2 v\Cdot p_1}\Big) \cr
    & = 
    -\frac{v\Cdot F_1 \Cdot F_2 \Cdot V_{12}\Cdot F_{3}\Cdot v}{ v\Cdot p_1 v\Cdot p_{12}} -  \frac{v\Cdot F_1 \Cdot F_3 \Cdot V_{1}\Cdot F_{2}\Cdot v}{ v\Cdot p_1 v\Cdot p_{13}} + \frac{v\Cdot F_1 \Cdot F_2 \Cdot F_3\Cdot v}{ v\Cdot p_1}\, , 
\end{split}
\end{align}
which indeed agrees precisely with the BCJ numerator $\npre([[1,2],3],v)$. Note that for this choice of the map $\langle\bullet\rangle_S$, any pre-numerator without leg $1$ at the beginning or end is set to zero. This explicitly shows why crossing symmetry is no longer manifest at the level of the pre-numerators.


\newpage

\bibliographystyle{utphys}
\bibliography{ScatEq}

\end{document}